\documentclass{jpp}
\usepackage{graphicx}
\usepackage{natbib}
\usepackage{float}
\usepackage[utf8]{inputenc}
\usepackage[T1]{fontenc}
\usepackage{xcolor}
\usepackage{amsmath,stackengine,scalerel,calrsfs}
\usepackage[hidelinks, linkcolor=blue, citecolor=blue, colorlinks=true]{hyperref}
\usepackage{enumitem}
\usepackage{bm}
\DeclareMathAlphabet{\pazocal}{OMS}{zplm}{m}{n}
\stackMath
\newcommand\reallywidehat[1]{%
\savestack{\tmpbox}{\stretchto{%
  \scaleto{%
    \scalerel*[\widthof{\ensuremath{#1}}]{\kern-.6pt\bigwedge\kern-.6pt}%
    {\rule[-\textheight/2]{1ex}{\textheight}}
  }{\textheight}%
}{0.5ex}}%
\stackon[1pt]{#1}{\tmpbox}%
}
\newcommand{\bb}{\textit{\textbf{b}}}

\newcommand{\bj}{\textit{\textbf{j}}}
\newcommand{\bk}{\textit{\textbf{k}}}
\newcommand{\bp}{\textit{\textbf{p}}}
\newcommand{\bq}{\textit{\textbf{q}}}
\newcommand{\bu}{\textit{\textbf{u}}}
\newcommand{\bx}{\textit{\textbf{x}}}
\newcommand{\by}{\textit{\textbf{y}}}
\newcommand{\bz}{\textit{\textbf{z}}}

\title{On local and non-local energy transfers in Hall magnetohydrodynamic turbulence}

\author{Arijit Halder\aff{1},
  Supratik Banerjee\aff{1}
  \corresp{\email{sbanerjee@iitk.ac.in}},
  Pablo D. Mininni,\aff{2}
 \and Manohar K. Sharma\aff{3}}

\affiliation{\aff{1}Department of Physics, Indian Institute of Technology, Kanpur 208016, India
\aff{2}Universidad de Buenos Aires, Facultad de Ciencias Exactas y Naturales, Departamento de Física, and CONICET - Universidad de Buenos Aires, Instituto de Física Interdisciplinaria y Aplicada (INFINA), Ciudad Universitaria, 1428 Buenos Aires, Argentina
\aff{3} Department of Applied Physics and Science Education
Eindhoven University of Technology, Eindhoven 5612AP, The Netherlands}

\begin{document}

\maketitle

\begin{abstract}
A systematic study of inertial energy cascade in three-dimensional incompressible Hall magnetohydrodynamic turbulence is conducted to probe into the locality of energy conserving triads and the subsequent transfers. Based on the nature of triadic conservations, the energy transfer due to the Hall term is further decomposed into two channels $\pazocal{BB}$ and $\pazocal{JB}$ corresponding to the terms $d_i (\bj\cdot\bnabla)\bb$ and $-d_i (\bb\cdot\bnabla)\bj$, respectively (\cite{Halder_2023, Banerjee_2024}). Here $\bb$ and $\bj$ represent the magnetic field and the current in Alfv\'en units, whereas $d_i$ is the ion inertial scale. Using direct numerical simulations, we calculate the shell-to-shell energy transfer rates corresponding to both the channels, and convincingly show each of them to comprise a combination of local and non-local energy transfers. A local inverse transfer is consistently observed at all scales of the channel $\pazocal{BB}$ whereas for the channel $\pazocal{JB}$, the local exchange of energy is associated with a gradual increase in strength as the scale is decreased, together with a transition from inverse to direct transfer across the Hall wavenumber, characterized by the ion inertial length $d_i$. Calculating mediator-specific transfer rates, we also conclude that a considerable amount of the local energy transfer is mediated by the non-local triads, especially at small scales of the channel $\pazocal{JB}$. The observed results can be explained using the power-law behaviour of the modal fields. The present study captures the intricate dynamics of energy transfer due to the Hall term and hence can be used to develop more insightful analytical models (shell models, for example) for Hall magnetohydrodynamic cascade. The framework can be extended to segregate the local and the nonlocal heating in various turbulent flows including ferrofluids, binary fluids, etc.
\end{abstract}

\section{Introduction}
Meticulous understanding of plasma turbulence is essential for explaining different nonlinear phenomena in space and astrophysics. At length scales larger than the ion inertial length ($d_i$), plasma behavior is described by magnetohydrodynamics (MHD), in which the dynamics of an ion-electron plasma is effectively captured by the center of mass fluid. In the limit of infinite conductivity, the magnetic field lines are frozen to the MHD fluid. However, for length scales comparable or inferior to $d_i$ but superior to the electron inertial scale  ($d_e$), the ion and electron fluids are no longer coupled and the magnetic field lines are frozen only in the electron fluid.  At those scales, one has to consider Hall MHD (HMHD) model where the induction equation is given by
\begin{equation}
\partial_t \bb - \eta\nabla^2 {\bb} =\bnabla\times(\bu\times\bb)- d_i \bnabla\times(\bj\times\bb), \label{ind_Hall}
\end{equation}
where $\bu$ is the velocity of the centre-of-mass fluid, $\bb$ is the magnetic field \textcolor{black}{expressed in Alfv\'{e}n units, $\eta$ is the magnetic diffusivity,} $\bj=\bnabla\times\bb$ represents the current and the last term on the \textit{r.h.s.} represents the Hall term. The presence of the Hall term distinguishes a HMHD flow from an ordinary MHD flow both in terms of the linear wave modes and turbulence. An ordinary incompressible MHD flow responds to a weak perturbation in terms of non-dispersive Alfv\'en waves, whereas the inclusion of the Hall term induces dispersion which mostly acts at scales comparable or smaller than $d_i$. For length scales much larger than $d_i$, one recovers the Alfv\'en waves. However, at scales much smaller than $d_i$, the dispersion relation splits into right-circularly polarized whistler and left-circularly polarized ion-cyclotron waves \citep{Lighthill_1960, Galtier_2016}.  Beyond the linear regime, HMHD turbulence distinguishes itself by the formation of fine scale current sheets and tubular structures \citep{Miura_2024}.  However, similar to MHD, the total energy $(\int \left(u^2+b^2\right)/2\ d\tau)$ is also an inviscid invariant of HMHD. In fully developed HMHD turbulence, one therefore expects a universal energy cascade with constant flux rate $\varepsilon$ across the extended inertial range comprising both MHD and Hall dominated sub-ion scales. In the realm of space and astrophysical plasmas, the cascade rate $\varepsilon$ is pivotal to unravel the mechanisms behind magnetic reconnection \citep{Huba_2004, Drake_2008, Sullivan_2009, Huang_2011}, turbulent heating \citep{Sorriso-Valvo_2007, Sahraoui_2009, MARINO_2023}, small-scale dynamos \citep{Mininni_2003, Mininni_2007, Kumar_2013} \textit{etc.} The analytical expressions of $\varepsilon$, obtained from exact relations in terms of two-point correlators or increments of relevant flow variables such as the velocity, the magnetic field \textit{etc.} \citep{Politano_1998, Galtier_2008, Banerjee_2013, Banerjee_2017, Andres_2018, Banerjee_2018, Andres_2019b}, are indeed used to calculate the turbulent heating rate
using both numerical and  observational data of plasma turbulence \citep{Hellinger_2018, Ferrand_2019, Andres_2019a, Bandyopadhyay_2020}. Interestingly, in Fourier space, the energy transfer induced by the Hall term is found to differ significantly from that of ordinary MHD. At MHD scales, the magnetic energy spectra follow a Kolmogorov-like $k^{-5/3}$ power law  whereas a  $k^{-7/3}$ spectrum emerges beyond the ion inertial scale \citep{Krishan_2004a, Krishan_2004b, Miura_2019}. In addition, for $k < k_H\ (\sim d_i^{-1})$, the modal transfer rate of magnetic energy due to the Hall term becomes negative, leading to a back-scatter of magnetic energy to the large scales \citep{Mininni_2007, Gomez_2010}.  To investigate the distinct nature of the energy transfers between MHD and HMHD in depth, it is customary to analyze in terms of the resonating energy conserving triads $(\bk,\bp,\bq)$ such that $\bk + \bp +\bq = \bm{0}$. The triads are said to be local if $k\simeq p\simeq q$ whereas for nonlocal triads we have one of the sides to be significantly smaller or larger than the other two sides (note that a side of the triangle can be arbitrarily small with respect to the other two sides, whereas it cannot be arbitrarily larger than the other two as mandated by the law of cosines).  On the other hand, the energy transfer is said to be local when the giver and the receiver modes are closely situated while the mediator mode, in principle, may be located anywhere. A local triad therefore always leads to local transfer whereas a nonlocal triad may associate a local or nonlocal transfer according as whether the giver and the receiver modes are of comparable sizes or not, respectively. 
Apart from small nonlocal transfers due to the energy containing large-scale structures in MHD \citep{ Alexakis_2005a, Carati_2006, Cho_2010}, the inertial scale energy transfers in both HD and MHD turbulence are predominantly mediated by the local triads \citep{Kraichnan_1971, Verma_2005b}. For HMHD turbulence, the situation becomes trickier. Whereas the total energy transfer remains local in nature, the energy transfer due to the Hall term shows clear signatures of nonlocality in the modal transfer of magnetic energy \citep{Mininni_2007, Gomez_2010, Araki_2011}. {Despite these studies}, a dedicated study to investigate the possible source of such nonlocality is still awaited. Also in those studies, the energy transfer was simply interpreted as an exchange between different modes of the $\bb$-fields only. However, $\bb$ and $\bj$ are connected with each other through space and time coordinates but one is not an explicit closed form of the other in the physical space. Whether all sorts of exchange due to the Hall term can be attributed to the $\bb$-field is a matter of debate. The total Hall term looks very similar to $\bnabla\times(\bu\times\bb)$, which is the nonlinear term in the induction of ordinary MHD. Albeit non-unique, one can be inspired by the  {$\bu$}-to-{$\bb$} and $\bb$-to-$\bb$  decomposition in ordinary MHD \citep{Plunian_2019} and decompose the Hall term into two parts as $\bnabla\times\left(\bj\times\bb\right)=-\left(\bj\cdot\bnabla\right)\bb + \left(\bb\cdot\bnabla\right)\bj$ to compare their individual contributions in the energy transfer\footnote{In the framework of HMHD, the contribution of the $\bj$-field is not explicitly present in the energy density. However, the two terms on the \textit{r.h.s} of Eq.~\eqref{ind_Hall} can be combined and decomposed as $\bnabla\times\left(\bu_e\times\bb\right)=\left(\bb\cdot\bnabla\right)\bu_e-\left(\bu_e\cdot\bnabla\right)\bb$ where $\bu_e = \bu - d_i \bj$ is the velocity of the electron fluid. In this way, the introduction of the Hall term induces an exchange between magnetic energy and electron fluid kinetic energy.}. The channel corresponding to $-\left(\bj\cdot\bnabla\right)\bb$ is labeled as channel $\pazocal{BB}$ and the channel corresponding to $\left(\bb\cdot\bnabla\right)\bj$ is labeled as channel $\pazocal{JB}$. In fact, rather than $\bj$, here the variable of interest is $-d_i\bj$, which is the relative velocity between electron and the ion fluids which essentially reduces to the electron fluid velocity in the small-scale limit $(kd_i\gg 1)$, where the motion of the ions can be neglected \citep{Biskamp_1999}.

Note that, this decomposition has already been used to explore the small-scale dynamo action and the mode-to-mode transfer rates due to the Hall term \citep{Halder_2023, Banerjee_2024}. By rigorous calculation, it was shown that the magnetic energy transfers through $\pazocal{BB}$ and $\pazocal{JB}$ channels are separately conserved in each triad $\bk + \bp +\bq =\bm{0}$. However, the nature of triadic conservation for the two channels is found to be structurally different \citep{Banerjee_2024}. The transfers through the channel $\pazocal{BB}$ are made up of three fundamental transfer units\footnote{In a triadic interaction, the fundamental transfer units constitute the base elements of the uniquely defined combined transfer rates $S(\bk|\bp, \bq)$, which denotes the net transfer rate of energy to the $\bk$-th mode from $\bp$ and $\bq$-modes with $\bk + \bp + \bq = \bm{0}$ (for a detailed discussion, see \cite{Banerjee_2024}).} whereas the transfers of the channel $\pazocal{JB}$ consist of five fundamental units. In addition, for the small-scale HMHD dynamos, the channel $\pazocal{JB}$ is found to be the primary contributor to the small scale turbulent Hall dynamo \citep{Halder_2023}. These aspects motivate a study to explore the direction and locality of energy transfer corresponding to each of the two channels.

In this paper, using direct numerical simulations (DNS) of HMHD turbulence with $512^3$ grid points, we thoroughly examine the locality (or nonlocality) of magnetic energy transfer through the aforesaid two channels. Employing shell-to-shell (S2S hereinafter) transfer rates, \textcolor{black}{we calculate the local and nonlocal contributions originating from each term, and identify the direct or inverse character of the resulting transfer.} While both the channels contain signature of back transfer to large-scales, their scale-wise behaviour are found to differ significantly from each other. In particular, for the channel $\pazocal{JB}$, a change of nature of the local transfer is observed across the Hall wavenumber $k_H$. To further characterize the locality of the triads, we compute S2S transfer rates for a fixed giver shell (denoted as the mediator-specific transfer later) and find a prominent disparity between the two channels. Finally, assuming a heuristic power-law behaviour of the modal fields, we propose some quantitative comparison on the basis of our observed results.

The paper is organized as follows: in Sec.~\ref{methods}, we present the governing equations, the definition of the energy transfer rates and discuss the numerical methods along with the simulation details. Sec.~\ref{Results} consists of calculation of flux and S2S transfer rates with and without fixing the giver shell and a heuristic quantitative interpretation of our findings. Finally, in Sec.~\ref{summary}, we summarize and conclude.   

\section{Methodology}
\label{methods}
\subsection{Governing equations and energy transfer rates}
\noindent The inviscid incompressible HMHD equations are given by 
\begin{align}
\partial_t \bu &= -(\bu\cdot \bm{\nabla})\bu +  (\bb\cdot \bm{\nabla})\bb -\bm{\nabla} P_T + \bm{f}, \label{eq:u1}\\
\partial_t \bb &=-(\bu\cdot \bm{\nabla})\bb +(\bb\cdot \bm{\nabla})\bu+ d_i\left(\bj\cdot\bnabla\right)\bb -d_i\left(\bb\cdot\bnabla\right)\bj, \label{eq:b1}\\
&\bm{\nabla} \cdot \bu = 0, \;\;  \bm{\nabla} \cdot \bb =  0, \label{eq:div}
\end{align}
where $P_T = p+b^2/2$ is the total pressure \textcolor{black}{and $\bm{f}$ is the external forcing.} To examine the locality of the energy conserving triads, it is customary to obtain the evolution equations for $n$-th shell energy densities 
$E^u_n$ and $E^b_n$ as 
\begin{align}
    \partial_t E^u_n &= \sum_m\partial_t E^u_{n,\ m} =\sum_m  \left[T^{u,\ u}_{n,\ m} +  T^{u,\ b}_{n,\ m}\right], \label{shell_Ek}\\
    \partial_t E^b_n &= \sum_m\partial_t E^b_{n,\ m} =\sum_m \left[ T^{b,\ b}_{n,\ m} + T^{b,\ u}_{n,\ m} +  \pazocal{H}^{b,\ b}_{n,\ m} + \pazocal{H}^{b,\ j}_{n,\ m}\right],\label{shell_Eb}
\end{align}
where $E^u_n\ =\sum_{|\bk|\in n}\bu_\bk\cdot\bu_\bk^* /2$ and $E^b_n\  =\sum_{|\bk|\in n}\bb_\bk\cdot\bb_\bk^* /2$, \textcolor{black}{with $\bu_\bk$ and $\bb_\bk$ denoting the Fourier transforms of $\bu(\bm{x})$ and $\bb(\bm{x})$, respectively. $T^{x,\ y}_{n,\ m}$ gives the net transfer rate from the field $\by_{\bp}$ to the field $\bx_{\bk}$, where $\left|\bp\right|$ and $\left|\bk\right|$ belong to the shells $m$ and shell $n$, respectively, and is mathematically expressed as}  
\begin{equation}
T^{x,\ y}_{n,\ m}= {\cal{Sgn}} (\bx, \by) \sum_{|\bk|\in n}\sum_{|\bp|\in m}^{\Delta}\Im\left[\left(\bp\cdot\bz_{\bq}\right)\left(\by_{\bp}\cdot\bx_{\bk}\right)\right]\label{S2S_generic}
\end{equation}
where $\bx$, $\by$ and $\bz$ can be $\bu$ and $\bb$ with $\Im$ denoting the imaginary part of a complex number, ${\cal{Sgn}} (\bx, \by) = 1$ if $\bx$ and $\by$ represent the same field and is equal to $-1$ otherwise, and $\Delta$ denotes the condition $\bk+\bp+\bq=\bm{0}$.
In the present context, the quantity $\Im\left[\left(\bp\cdot\bz_{\bq}\right)\left(\by_{\bp}\cdot\bx_{\bk}\right)\right]$ represents the field specific mode-to-mode transfer rate from $\by_\bp$ to $\bx_\bk$, with $\bk$, $\bp$ and $\bq$ being denoted as the receiver, giver and mediator wave modes, respectively \citep{DAR_2001, Verma_2004, Banerjee_2024}. Note that, these transfer rates
are defined without symmetrizing the triads and the definitions of giver and mediators can be debatable. However, previous works have shown that
shell-to-shell transfer rates can be properly defined (\textit{i.e.}, they conserve energy and
have all properties expected for a transfer function), and calculated much more easily if one uses the non-symmetrized expressions \citep{DAR_2001, Teimurazov_2018}.Nonetheless, it is important to recognize that a phenomenological interpretation based on the concepts of the giver and mediator wavenumbers constitutes just one of several plausible interpretations.  Similar to the aforementioned nonlinear transfers, $\pazocal{H}^{b,\ b}_{n,\ m}$ and $\pazocal{H}^{b,\ j}_{n,\ m}$ represent the transfer rates associated with $d_i\left(\bj\cdot\bnabla\right)\bb$ and $-d_i\left(\bb\cdot\bnabla\right)\bj$, respectively and are written as 
\begin{align}
\pazocal{H}^{b,\ b}_{n,\ m}&=-d_i\sum_{|\bk|\in n}\sum_{|\bp|\in m}^{\Delta}\Im\left[\left(\bp\cdot\bj_{\bq}\right)\left(\bb_{\bp}\cdot\bb_{\bk}\right)\right],\label{S2S_bb}\\
\pazocal{H}^{b,\ j}_{n,\ m}&=d_i\sum_{|\bk|\in n}\sum_{|\bp|\in m}^{\Delta}\Im\left[\left(\bp\cdot\bb_{\bq}\right)\left(\bj_{\bp}\cdot\bb_{\bk}\right)\right].\label{S2S_bj}
\end{align}
Following the standard nomenclature of mode-to-mode transfers, it is straightforward to recognize the transfers in Eqs.~\eqref{S2S_bb} and \eqref{S2S_bj} to be associated with the channels $\pazocal{BB}$ and $\pazocal{JB}$, respectively. In HMHD, however, no term of energy transfer is obtained by taking the scalar product of the HMHD equations with $\bj$ and hence the back transfer for the channel $\pazocal{JB}$ does not appear. In the shell picture, the transfer is said to be local if the shell number $n$ mostly receives energy from the neighboring shells whereas for the nonlocal transfers a non-negligible contribution also comes from the distant shells.

\subsection{Numerical details}
\label{Numericals}
\begin{figure}
    \centering
    \includegraphics[width=\columnwidth]{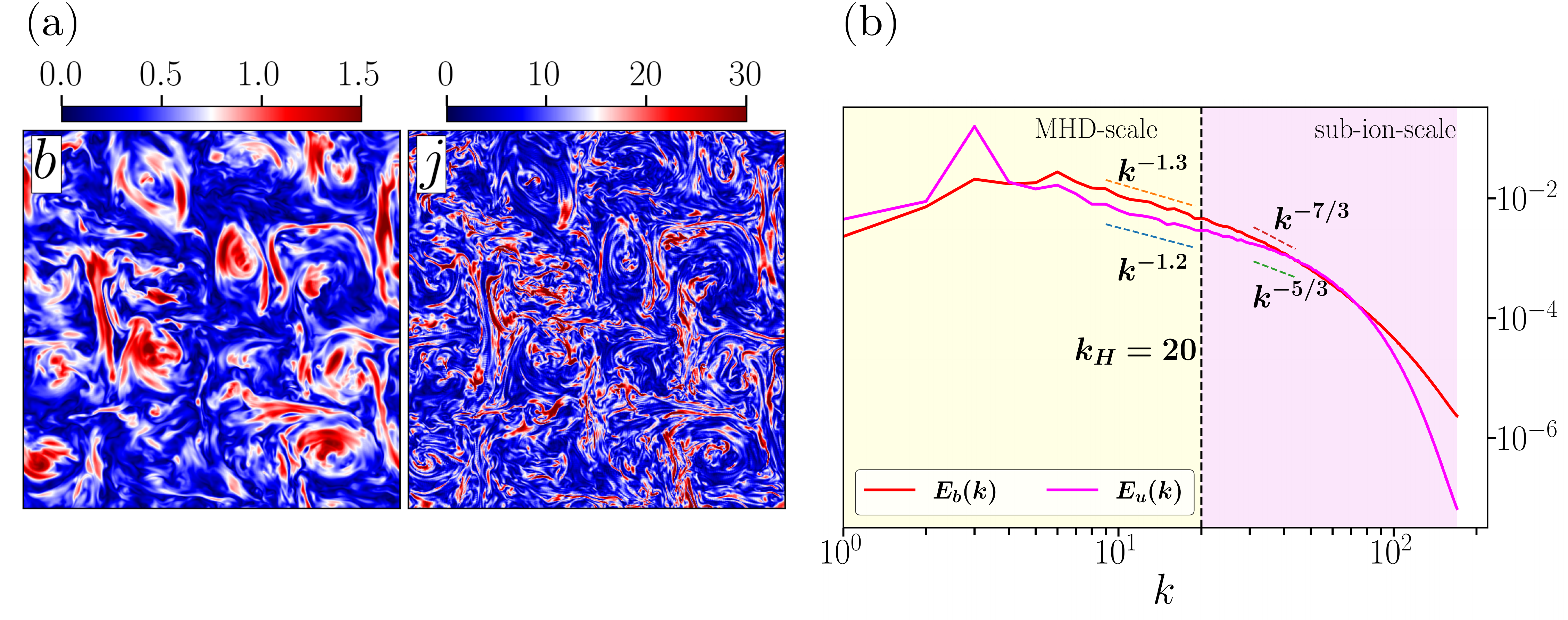}
    \caption{(a) Instantaneous slice of absolute values of magnetic and current fields at statistical steady state. Both fields show signature of a fully developed turbulent flow. \textcolor{black}{(b) Kinetic and magnetic energy spectra at steady state as a function of wavenumber $k$. The dashed vertical line represents the Hall wavenumber $k_H$.} }
    \label{fig:b_j_slice}
\end{figure}
\setlength{\tabcolsep}{0.5em}
{\renewcommand{\arraystretch}{1.5}
\begin{table}
  \begin{center}
\def~{\hphantom{0}}
  \begin{tabular}{cccccc}
      $N$  & $\nu_h=\eta_h$ & $d_i$ & $f_0$     & $k_{d}$ & $k_{max}/k_d$ \\
      512  & $10^{-7}$ & $0.05$ & $0.5$      & 100 & 1.7 \\
  \end{tabular}
  \caption{\textcolor{black}{Simulation parameters. $N^3$ is the number of grid points, $\nu_h$ is the hyperviscous coefficient, $\eta_h$ is the hyperdiffusive coefficient, $d_i$ is the ion inertial length, $f_0$ is the external forcing amplitude and $k_{d}$ is the hyperviscous Kolmogorov wavenumber.}  }
  \label{tab:parameters}
  \end{center}
\end{table}}

We solve \eqref{eq:u1} and \eqref{eq:b1} using a parallel pseudo-spectral code in a 3D periodic domain with $512^3$ grid points. The box length is taken to be $2\pi$ in each spatial direction. In our simulation, we take $d_i =0.05$ beyond which the Hall effect is expected to be visible. The aliasing error is removed with a standard 2/3-dealiasing method which fixes the maximum available wavenumber $k_{max}\approx 171$ \citep{Orszag_1971}. A fourth-order Runge-Kutta (RK4) scheme is used for time integration and the time step is adjusted with CFL condition. The turbulence is generated and sustained by Taylor-Green (TG) forcing $\bm{f}= f_0 [sin(k_0x)cos(k_0y)cos(k_0z), -cos(k_0x)sin(k_0y)cos(k_0z)]$ with $f_0 = 0.5$ and $k_0=2$ which corresponds to the forcing wavenumber $k_f = k_0\sqrt{3}\approx 3.4$. A clear inertial range can be obtained if the dissipative terms contain higher order derivatives as compared to the nonlinear terms. In order to avoid the contamination of Hall contribution due to the usual dissipative effects $(\sim \nabla^2)$, here we employ a $\nabla^4$-hyperdiffusive operator in both the momentum and induction equation. Note that, the magnetic Prandtl number for our simulation is taken to be unity. The simulation is well-resolved as the ratio $k_{max}/k_d > 1$ where $k_d = (\varepsilon/\nu_h^3)^{(1/10)}$ is the hyperviscous Kolmogorov wavenumber where $\varepsilon$ is the total dissipation rate and $\nu_h$ is the hyperviscous coefficient \citep{Banerjee_2025}. The relevant simulation parameters are summarized in Table~\ref{tab:parameters}. All the numerical calculations are performed when the system reaches a statistical steady state balancing the average injection and dissipation of total energy.  An instantaneous slice of the magnetic and current fields (see Fig.~\ref{fig:b_j_slice}a) shows the clear signature of fully-developed homogeneous turbulence. In Fig.~\ref{fig:b_j_slice}b, we plot the power spectra for magnetic and kinetic energies as a function of wavenumber $k$. For scales larger than the $d_i$, magnetic and kinetic energies follow $k^{-1.3}$ and $k^{-1.2}$, respectively whereas for scales below $d_i$, the magnetic and the kinetic energy spectra follow $k^{-7/3}$ and $k^{-5/3}$ power-laws, respectively. The spectra for scales larger than $d_i$ are shallower than the ones reported in earlier studies on HMHD turbulence \citep{Miura_2019, Yadav_2022}. In our simulations with $512^3$ grid points, the forcing wavenumber $k_f \approx 3.4$ and the Hall wavenumber $k_H\approx 20$ are close and hence a well-separated MHD inertial range is not obtained. However, an extended inertial range with kinetic and magnetic spectral exponents close to $-1.5$ and $-1.6$ is obtained for a simulation with $512^3$ grid points and $d_i = 0.02$, corresponding to $k_H \approx 50$ (see Fig. \ref{fig:spectra_flux_kH50}a in Sec. \ref{effect_kH}). Nevertheless, we continue using $k_H = 20$ throughout the entire analysis in order to keep the Hall wavenumber approximately in the middle of the logarithmic wavenumber range (see Fig.~\ref{fig:b_j_slice}a). One may then question the robustness of the results as $k_H$ changes. To verify that, we have explicitly calculated the S2S transfer rates and show that the main conclusion regarding the nature of energy transfer in the two channels remains unchanged (see Figs. \ref{fig:S2S_kH50}a and b in Sec. \ref{effect_kH}).

To capture a clear picture of the energy transfer in the inertial range, we construct $25$ concentric shells with the radii: $k_n\in \{0, 2, 4, 8, ..., 128, 256\}$ where the 
radii between $8$ and $128$ is logarithmically binned with the ratio $k_n/k_{n-1}=1.15$, between the subsequent shells. Note that, in our case, the inner and outer radii of the $n$-th shell are $k_{n-1}$ and $k_{n}$, respectively. The choice of logarithmic binning for the intermediate wavenumbers is adopted to ensure localization of small turbulent eddies in the physical space \citep{Eyink_2009}. This is due to the fact that shell thickness grows with increasing radius, enabling the accumulation of large number of Fourier modes in the high wavenumber region. In contrast, linear binning generates shells of uniform thickness which does not necessarily guarantee such localization (see Appendix \ref{appA}). Using the aforementioned radii, we also calculate the corresponding flux rates for kinetic energy, magnetic energy (MHD part), the Hall term, the total energy injection and dissipation as 
\begin{align}
\Pi_u (k_0) =&   -\sum_{|\bk|\leq k_0} \sum_{|\bp|> k_0}^{\Delta}\Im\left[ (\bp\cdot\bu_\bq)(\bu_\bp\cdot\bu_\bk)-(\bp\cdot\bb_\bq)(\bb_\bp\cdot\bu_\bk)\right],\label{Pi_u}\\
\Pi_b (k_0) =&  - \sum_{|\bk|\leq k_0}\sum_{|\bp|> k_0}^{\Delta} \Im\left[ (\bp\cdot\bu_\bq)(\bb_\bp\cdot\bb_\bk)-(\bp\cdot\bb_\bq)(\bu_\bp\cdot\bb_\bk)\right],\label{Pi_b}\\
\Pi_{Hall} (k_0) =& -d_i  \sum_{|\bk|\leq k_0}\sum_{|\bp|> k_0}^{\Delta} \Im\left[ (\bp\cdot\bb_\bq)(\bj_\bp\cdot\bb_\bk)-(\bp\cdot\bj_\bq)(\bb_\bp\cdot\bb_\bk)\right]\nonumber\\
& - d_i  \sum_{|\bk|\leq k_0}\sum_{|\bp|\leq  k_0}^{\Delta}\Im\left[ (\bp\cdot\bb_\bq)(\bj_\bp\cdot\bb_\bk)\right],\label{Pi_H}\\
\Pi_{inj} (k_0) =& \sum_{|\bp|\leq k_0} \Re\left[\bu_{\bp}^*\cdot \bm{f}_{\bp}\right]\;\; \text{and}  \\
 \Pi_{D} (k_0) =& -\sum_{|\bp|\leq k_0} \left[ \nu_h p^{4} |\bu_{\bp}^*|^2  + \eta_h p^{4} |\bb_{\bp}^*|^2  \right],
\end{align}
respectively\footnote{The expression for flux rates in Eqs.\eqref{Pi_u}-\eqref{Pi_H} are based on the mode-to-mode formalism outlined above and are equivalent to the definition provided in \cite{Kraichnan_1959}. A formal proof of this equivalence is demonstrated in \citet{Plunian_2019}.}.

It is interesting to note that, the term $- d_i  \sum_{|\bk|\leq k_0}\sum_{|\bp|\leq  k_0}^{\Delta}\Im\left[ (\bp\cdot\bb_\bq)(\bj_\bp\cdot\bb_\bk)\right]$ in Eq.~\eqref{Pi_H} represents the net transfer rate of magnetic energy with both $|\bp|$ and $|\bk|$ being inside a sphere of radius $k_0$. For transfer of energy from $\bm{x}$-to-$\bm{y}$ field, this particular term gets cancelled with similar type of contribution emerging from the corresponding back transfer from $\bm{y}$-to-$\bm{x}$. This is true for $\bu$-to-$\bu$, $\bb$-to-$\bb$ and $\bb$-to-$\bu$ transfers where the corresponding back transfers automatically exist for the first two cases and also holds for the third case where $\bu$-to-$\bb$ transfers participate in the resulting energy cascade \citep{Verma_2019, Banerjee_2024}. In case of HMHD, the energy transfer involves channels $\pazocal{BB}$ and $\pazocal{JB}$ where the back transfer for the channel $\pazocal{JB}$ does not appear in the energy transfer. The aforesaid contribution is therefore not counter-balanced and survives in the calculation of $\Pi_{Hall}$ (see Appendix \ref{appB}).

\section{Results and Discussion}
\label{Results}
\subsection{Behaviour of flux rates}
\label{flux_numerics}
\begin{figure}
    \centering
    \includegraphics[width=\columnwidth]{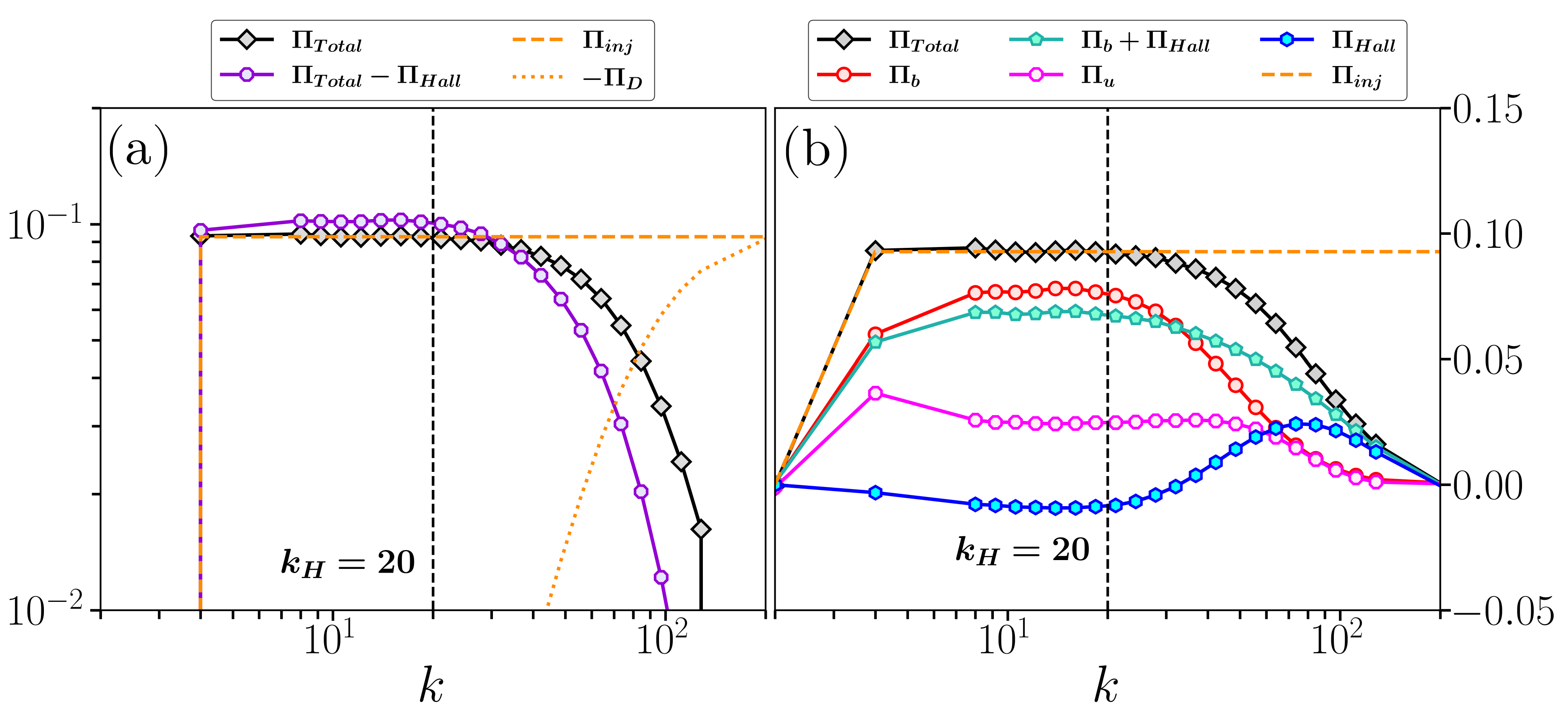}
    \caption{(a) Total energy flux rates, with and without the contribution from the Hall term, as a function of wavenumber $k$. $\Pi_{inj}$ and $\Pi_{D}$ respectively denote the flux rates due to injection and total energy dissipation. (b) Various components of the total flux as function of wavenumber $k$. Here, $\Pi_{b}$, $\Pi_{u}$, and $\Pi_{Hall}$ respectively denote flux rates for magnetic, kinetic energies and due to the Hall term with $\Pi_{Total} = \Pi_{b}+\Pi_{u}+\Pi_{Hall}$. }
    \label{fig:flux_comps_spectra}
\end{figure}

In Fig.~\ref{fig:flux_comps_spectra}a, we plot the total energy flux $\Pi_{Total}\ (=\Pi_u + \Pi_b + \Pi_{Hall})$ as a function of wavenumber $k$, using the expressions provided in Eqs.\eqref{Pi_u}-\eqref{Pi_H}. For the sake of comparison and to delineate the role of the Hall term, we plot $\Pi_{Total}$ with and without $\Pi_{Hall}$. As it is clearly observed, $\Pi_{Total}$ closely follows the injection flux rate $\Pi_{inj}$ (dashed orange curve in Fig.~\ref{fig:flux_comps_spectra}a) across a range of intermediate wavenumbers, indicating a Kolmogorov-like universal regime for the total energy cascade. For clarity, we also plot the dissipation flux rate $-\Pi_{D}$ (dotted orange curve in Fig.~\ref{fig:flux_comps_spectra}a). Evidently, $-\Pi_{D}$ remains negligible compared to $\Pi_{Total}$ up to the hyperviscous Kolmogorov wavenumber $k_d$, where it eventually takes over, ensuring an approximately dissipation-free inertial cascade regime. To further probe into the behaviour of various components of $\Pi_{Total}$, we plot $\Pi_u$, $\Pi_b$ and $\Pi_{Hall}$ as a function of $k$ in Fig.~\ref{fig:flux_comps_spectra}b. For length scales greater than $d_i$, i.e., $k d_i<1$, the magnetic energy transfer rate $\Pi_b$ dominates over $\Pi_u$ and $\Pi_{Hall}$. Similar to previous studies \citep{Mininni_2007, Gomez_2010, Halder_2023}, $\Pi_{Hall}$ becomes slightly negative across the said range of scales. However, for $k d_i >1$, $\Pi_{Hall}$ becomes positive and gradually exceeds $\Pi_b$ and $\Pi_u$ at sufficiently small scales. As a result, for $kd_i < 1$, the net magnetic energy transfer rate (MHD + Hall) decreases towards the small-scale, whereas for $kd_i > 1$, it becomes amplified. Interestingly, both $\Pi_u$ (pink curve in Fig.~\ref{fig:flux_comps_spectra}b) and $\Pi_b + \Pi_{Hall}$ (green curve in Fig.~\ref{fig:flux_comps_spectra}b) remain approximately constant across the inertial range, suggesting a decoupled cascade of kinetic and magnetic energies. In ordinary MHD, such decoupling is attributed to the suppression of magnetic field-line stretching in the inertial regime \citep{Bian_2019}. However, for HMHD, a similar type of reasoning requires further investigation.

The flux rates discussed so far only provide the net transfer rate to a particular mode from all the others. In order to get a more detailed picture of the energy cascade and to investigate its locality, it is appropriate to study the corresponding S2S transfer rates using Eq.~\eqref{S2S_generic}. 

\subsection{Calculation of shell-to-shell transfer rates}
\label{S2S_numerics}

\begin{figure}
    \centering
    \includegraphics[width=0.7\columnwidth]{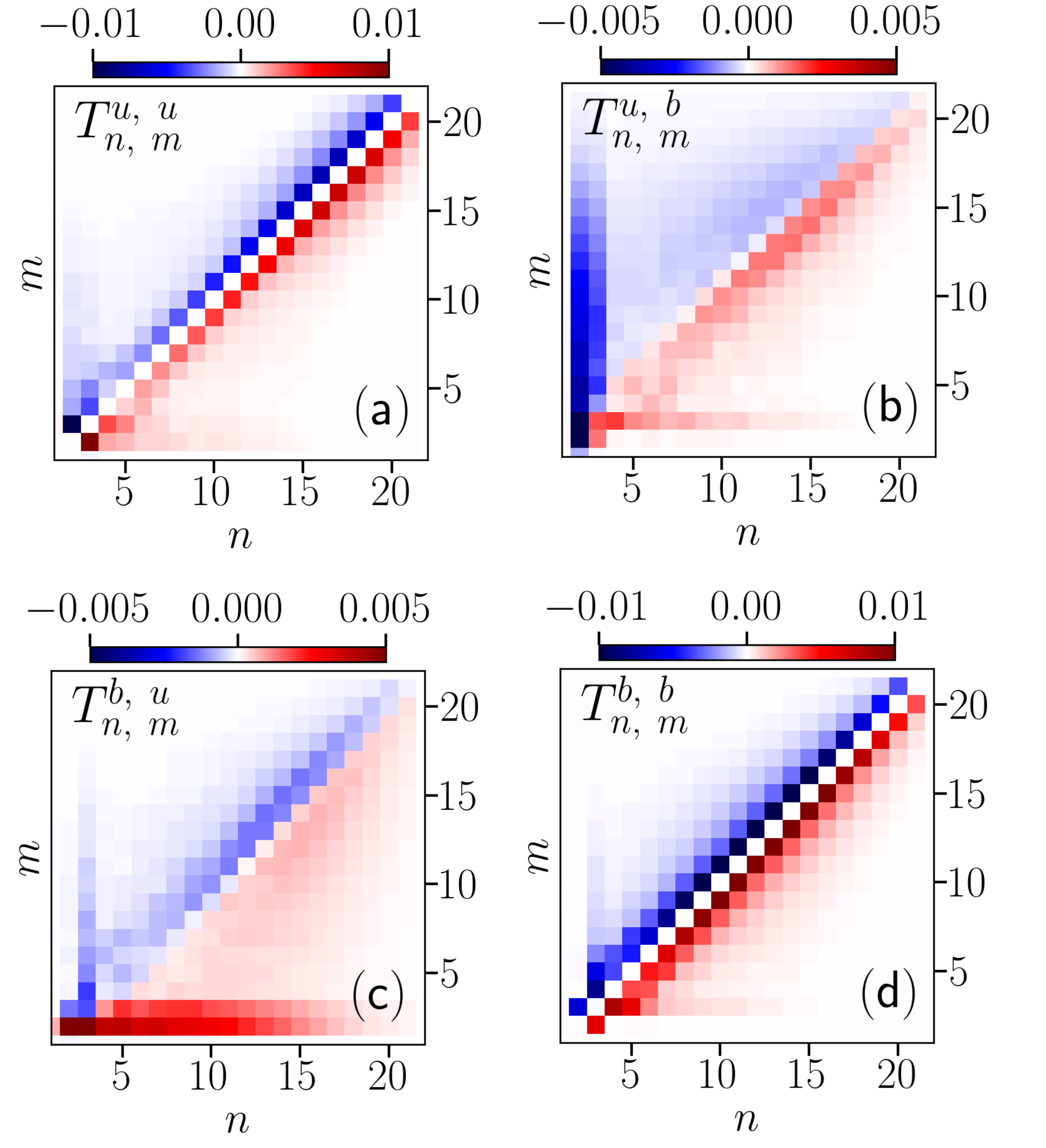}
    \caption{S2S transfer rates corresponding to MHD nonlinear terms as a function of giver and receiver shells $(m, n)$. $T^{u,\ u}_{n,\ m}$ and $T^{u,\ b}_{n,\ m}$ denote the transfer rate of kinetic energy through $\bu$-to-$\bu$, $\bb$-to-$\bu$ channels respectively. $T^{b,\ u}_{n,\ m}$ and $T^{b,\ b}_{n,\ m}$ denote the transfers rates of magnetic energy through $\bu$-to-$\bb$ and $\bb$-to-$\bb$ channels respectively. }
    \label{S2S_MHD_Combined}
\end{figure}

First, we examine $T^{u,\ u}_{n,\ m}$ and $T^{u,\ b}_{n,\ m}$ which denote the S2S transfer rates of kinetic energy through the $\bu$-to-$\bu$ and $\bb$-to-$\bu$ channels, respectively. For a given value of the receiver shell number $n$, if $T^{x,\ y}_{n,\ m}$ is positive for $n>m$ or negative for $n<m$, then we have a direct transfer and vice versa. From Fig.~\ref{S2S_MHD_Combined}, both $T^{u,\ u}_{n,\ m}$ and $T^{b,\ b}_{n,\ m}$ represent a direct \textcolor{black}{transfer} which is mainly local in nature \textit{i.e.} for a receiver shell number $n$, significant contribution only comes from the neighboring shells. In contrast, for both cases of cross-transfer $T^{u,\ b}_{n,\ m}$ and $T^{b,\ u}_{n,\ m}$, a non-negligible nonlocal transfer is observed involving the scale of energy injection. This agrees with the previous findings of nonlocal transfers in MHD due to the effect of large-scale energy-containing structures \citep{Alexakis_2005a, Carati_2006, Cho_2010}.

\begin{figure}
    \centering
    \includegraphics[width=\linewidth]{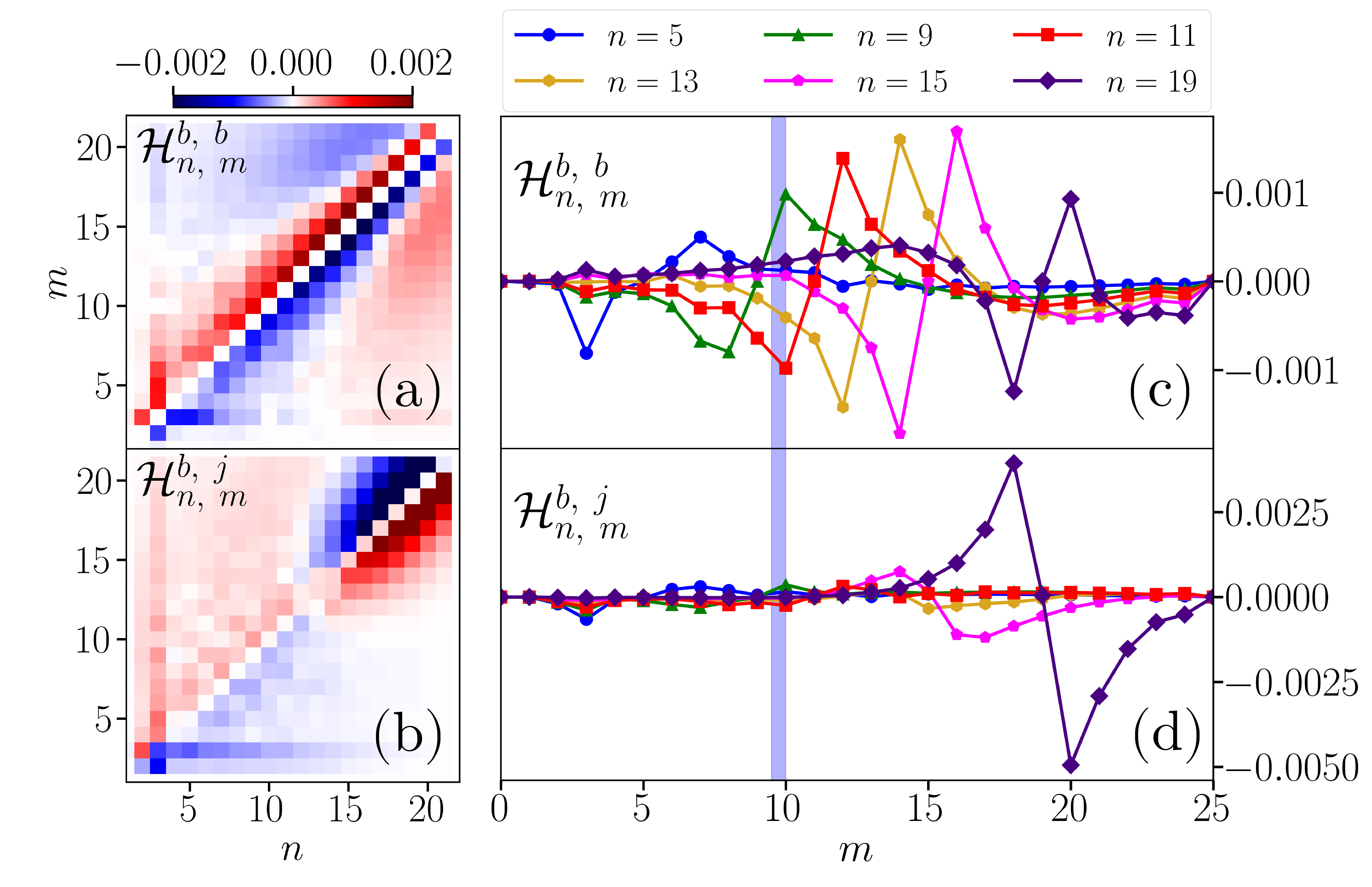}
    \caption{(a) and (c) S2S transfer rates for the channel $\pazocal{BB}$ of the Hall term as a function of giver and receiver shell numbers ($m$, $n$). (b) and (d) Same for the channel $\pazocal{JB}$ as a function of giver and receiver shell numbers ($m$, $n$). }
    \label{S2S_HALL_TWO_CHANNELS}
\end{figure}

\begin{figure}
    \centering
    \includegraphics[width=0.8\columnwidth]{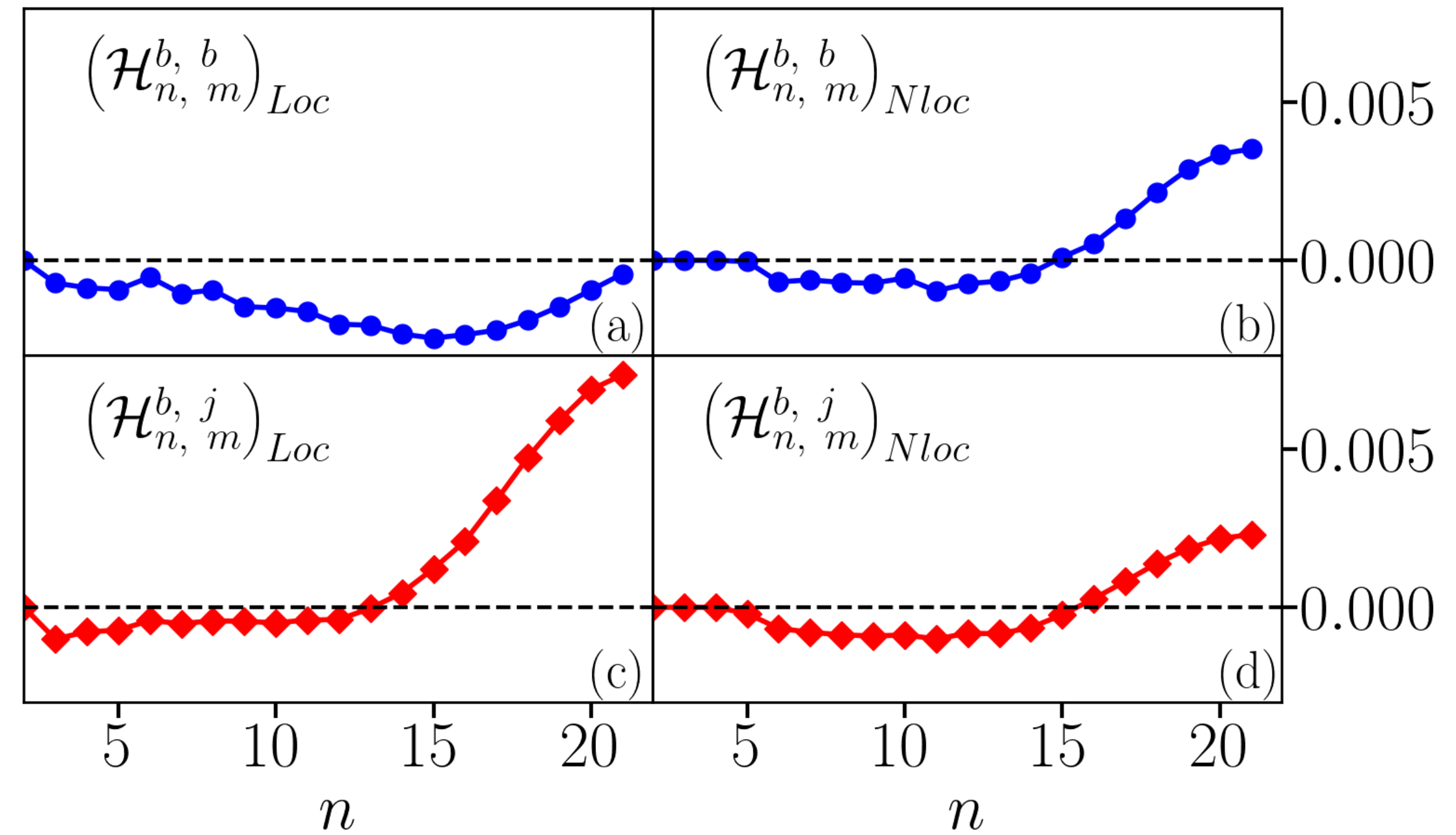}
    \caption{(a) and (b) Local and nonlocal magnetic energy transfer rates due to $\pazocal{H}^{b,\ b}_{n,\ m}$ (plotted in blue color) as a function of $n$, respectively. (c) and (d) Similar plots for $\pazocal{H}^{b,\ j}_{n,\ m}$ (in red color). For a given $n$, the local transfers rates are computed only for $m=n-1$ and $n-2$.  }
    \label{Two_channel_local_flux}
\end{figure}

In Figs.~\ref{S2S_HALL_TWO_CHANNELS}a and b, we plot the S2S transfer rates corresponding to channels $\pazocal{BB}$ and $\pazocal{JB}$ of the Hall term as a function of giver and receiver shell numbers $(m,\  n)$, respectively. As it is apparent, both the channels are comprised of local and nonlocal transfers.
For $\pazocal{H}^{b,\ b}_{n,\ m}$, close to the diagonal, for a fixed giver shell $m$, positive transfer is observed for a receiver shell $n<m$ (red squares above the diagonal) whereas negative transfer is observed for a receiver shell $n>m$ (blue squares below the diagonal). This clearly signifies a local inverse transfer of magnetic energy. On the other hand, far from the diagonal, the nature of the transfer is reversed and therefore a negative transfer is found above the diagonal while a positive transfer is found below the diagonal, indicating a direct nonlocal transfer of magnetic energy. Interestingly, the strength of the nonlocal transfers becomes gradually important at small scales \textit{i.e.} for large values of $m$ and $n$ (see Fig.~\ref{S2S_HALL_TWO_CHANNELS}a). The nature of the energy transfer due to $\pazocal{H}^{b,\ j}_{n,\ m}$, on the other hand, is remarkably different (see Fig.~\ref{S2S_HALL_TWO_CHANNELS}b). For shells with $m,n < 12$, both the local and nonlocal transfer rates of $\pazocal{H}^{b,\ j}_{n,\ m}$ are found to be relatively weaker compared to those of $\pazocal{H}^{b,\ b}_{n,\ m}$ whereas the transfer rate considerably enhances at smaller scales where it is principally nourished by the local S2S transfers. Furthermore, the color code of the transfers clearly indicates an inverse transfer of energy for small wavenumbers whereas the transfer becomes direct at sufficiently small scales. Interestingly, in both channels, a considerable transfer associating $m,\ n = 3$ is observed which is possibly induced by the interaction between the injection and the Hall wavenumbers.  
\setlength{\tabcolsep}{0.5em}
{\renewcommand{\arraystretch}{1.5}
\begin{table}
  \begin{center}
\def~{\hphantom{0}}
  \begin{tabular}{cc}
      Shell no. $(n)$  & Wavenumber range  \\
      5  & $[9.19,\ 10.56]$\\
      9  & $[16.00,\ 18.38]$\\
      10 & $[18.38,\ 21.11]$\\
      11 & $[21.11,\ 24.25]$\\
      13 & $[27.86,\ 32.00]$\\
      15 & $[36.76,\ 42.22]$\\
      19 & $[64.00,\ 73.52]$\\
  \end{tabular}
  \caption{Receiver shell indices and the corresponding wavenumber ranges. }
  \label{tab:shell_indices}
  \end{center}
\end{table}}

To further compare the strength of local and nonlocal transfer rates for the two channels of the Hall term, we plot $\pazocal{H}^{b,\ b}_{n,\ m}$ and $\pazocal{H}^{b,\ j}_{n,\ m}$ as function of $m$ for $n=5,\ 9,\ 11,\ 13,\ 15$ and $19$, respectively (see Figs.~\ref{S2S_HALL_TWO_CHANNELS}c and d). The corresponding wavenumber ranges are summarized in Table \ref{tab:shell_indices}. Such a diagram is useful to recognize the principal contributors for a fixed receiver shell $n$.
For every single $n$, the peaks of the transfer are situated at neighboring shells indicating a predominant local transfer for $\pazocal{H}^{b,\ b}_{n,\ m}$. In addition, the amplitudes of the local transfers increase with receiver shell number $n$ running from $5$ to $15$. A similar trend is also observed for the negative transfers at large $m$. These two features offer an alternative way to delineate the increasing strength of both the local and nonlocal transfers at small scales.  
For $n = 19$, however, the amplitude for the local transfer is found to decrease along with a substantial nonlocal contribution from small values of $m$. For $\pazocal{H}^{b,\ j}_{n,\ m}$, the situation is notably different. For $n<15$, we observe a comparatively weaker inverse transfer made up of comparable contributions both from local and nonlocal giver shells. For $n = 15$ and beyond, the direction of the energy transfer flips and local contributions become quite large with respect to the nonlocal contributions. Interestingly, similar to the $\pazocal{H}^{b,\ b}_{n,\ m}$, here also a considerable (both local and nonlocal) transfer is observed around $m,\ n = 3$.

In order to exclusively capture the contributions from the local shells, we calculate the total magnetic energy flux rate up to second nearest neighbour shell (see Figs.~\ref{Two_channel_local_flux}a and c). As an example, for receiver shell $n$, the total flux rate is computed for giver shells $m = n-1$ and $m = n-2$. 
Note that there is no standard threshold value to distinguish between the local and the nonlocal transfers. For the present study, a rough idea about the extent of the local transfers can be obtained from the S2S transfers in MHD (see Fig.~\ref{S2S_MHD_Combined}). As it is apparent, for a given receiver shell $n$, the transfers die down beyond $m = n\pm 2$. In our work, this is taken as a benchmark for designating local and nonlocal transfers. In fact, from Figs.~\ref{S2S_HALL_TWO_CHANNELS}c and d, one can see that even for the Hall term such a choice is justified as the nearby contributions to a shell $n$ mostly comes from $m = n\pm 2$ before the first fall of the transfer rates. As expected, for lower wavenumbers, the local contribution for the channel $\pazocal{BB}$ is stronger compared to the channel $\pazocal{JB}$. Around shell number 15, the local contribution of the latter sharply increases and becomes positive signifying direct S2S transfer. Evidently, the local contributions for the channel $\pazocal{JB}$ is primarily responsible for the observed enhancement of magnetic energy transfer towards smaller scales discussed previously. In contrast, the nonlocal contributions for both the channels $\pazocal{BB}$ and $\pazocal{JB}$ are inverse and of comparable magnitudes for shells below $ n = 15$. Beyond that, both become direct with $\pazocal{H}^{b,\ b}_{n,\ m}$ being slightly greater than $\pazocal{H}^{b,\ j}_{n,\ m}$ (see Figs.~\ref{Two_channel_local_flux}b and d). 

\begin{figure}
    \centering
   \includegraphics[width=0.4\linewidth]{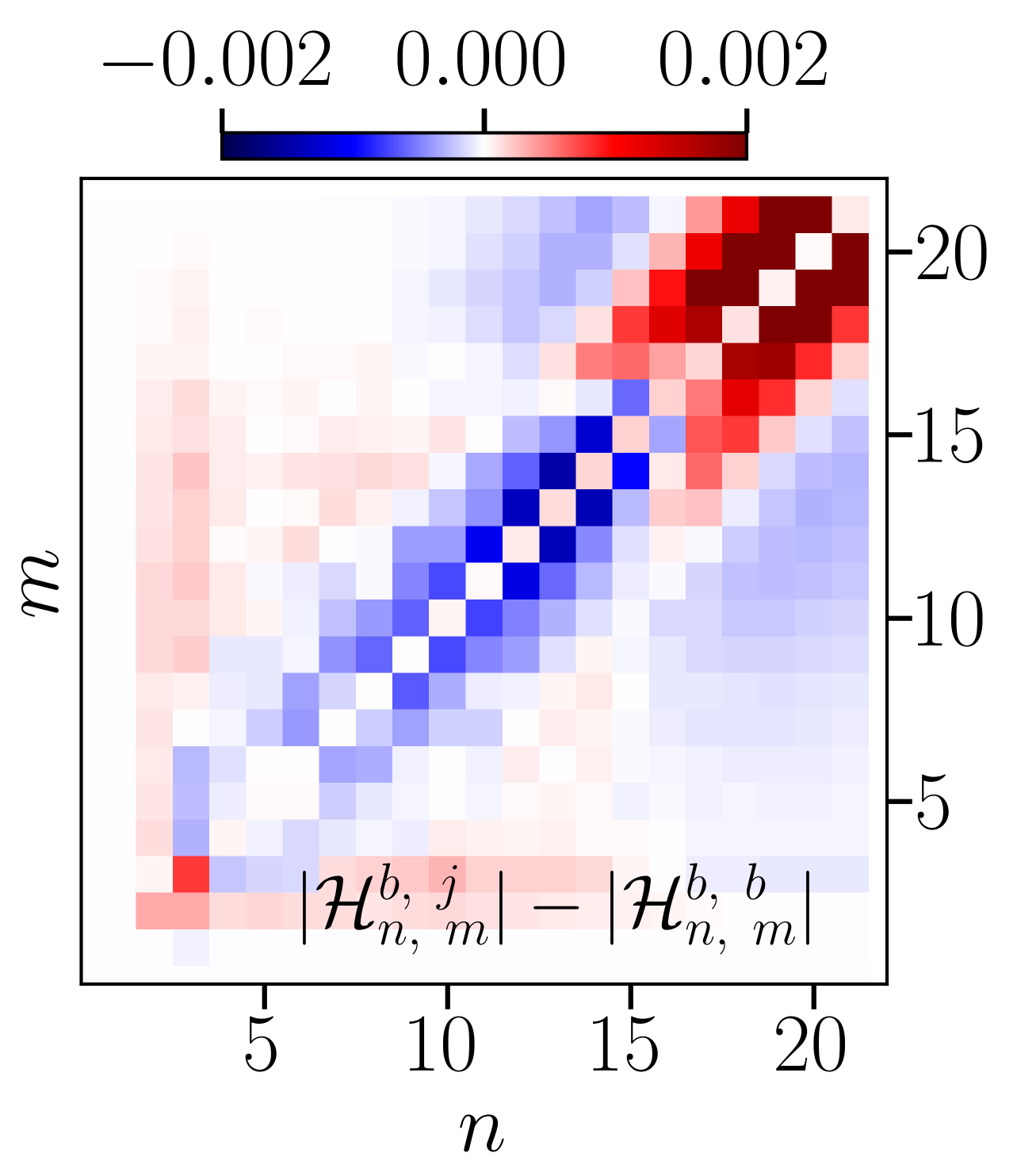}
    \caption{$|\pazocal{H}^{b,\ j}_{n,\ m}| - |\pazocal{H}^{b,\ b}_{n,\ m}|$ as a function of giver-receiver shell numbers $(m, n)$.}
    \label{fig:diff_abs_SBJ_minus_abs_SBB}
\end{figure}
Finally, to summarize the relative strength of the two channels of the Hall term, in Fig.~\ref{fig:diff_abs_SBJ_minus_abs_SBB}, we plot $|\pazocal{H}^{b,\ j}_{n,\ m}| - |\pazocal{H}^{b,\ b}_{n,\ m}|$ as a function of $(m,\ n)$. For a given $(m,\ n)$, positive and negative values of the transfer rate respectively denote the dominance of $|\pazocal{H}^{b,\ j}_{n,\ m}|$ and $|\pazocal{H}^{b,\ b}_{n,\ m}|$, respectively. As it is apparent, local transfers in smaller scales (high values of $m$ and $n$) are dominated by $\pazocal{H}^{b,\ j}_{n,\ m}$ whereas large-scale (low values of $m$ and $n$) local transfers are primarily governed by $\pazocal{H}^{b,\ b}_{n,\ m}$. For the nonlocal transfers, on the other hand, the regions with $m > n$ is dominated by $\pazocal{H}^{b,\ j}_{n,\ m}$ whereas $\pazocal{H}^{b,\ b}_{n,\ m}$ becomes important for $m < n$. 
A possible insight into the different nature of the transfer for the two channels can be obtained by expressing the transfers in terms of two-point correlators in physical space. To achieve that, we start with the two-point magnetic energy correlator $\mathcal{R}_b = \langle \bb\cdot\bb' \rangle/2$ where primed and unprimed quantities denote variables at positions $\bx + \bm{r}$ and $\bx$, respectively and $\langle(\cdot)\rangle$ denotes statistical average (equivalent to space average for homogeneous turbulence). Using Eq.~\eqref{eq:b1} and considering the Hall contribution only, one can write 
\begin{equation}
    \partial_t \mathcal{R}_b = \mathcal{T}_{b\rightarrow b}^{H} + \mathcal{T}_{j\rightarrow b}^{H},
\end{equation}
where 
\begin{align}
\mathcal{T}_{b\rightarrow b}^{H} &= \frac{d_i}{2}\left\langle\bb\cdot\left[\left(\bj'\cdot\bnabla'\right)\bb'\right]+\bb'\cdot\left[\left(\bj\cdot\bnabla\right)\bb\right]\right\rangle\;\; \text{and}\label{eq:Tbb}\\
\mathcal{T}_{j\rightarrow b}^{H} &= \frac{d_i}{2}\left\langle-\bb\cdot\left[\left(\bb'\cdot\bnabla'\right)\bj'\right]-\bb'\cdot\left[\left(\bb\cdot\bnabla\right)\bj\right]\right\rangle\nonumber\\ 
&= \frac{d_i}{4} \left\langle -\bb'\cdot\left[\left(\bb\cdot\bnabla\right)\bj\right]-\bb\cdot\left[\left(\bb'\cdot\bnabla'\right)\bj'\right]
-\bj'\cdot\left[\left(\bb\cdot\bnabla\right)\bb\right]-\bj\cdot\left[\left(\bb'\cdot\bnabla'\right)\bb'\right]\right\rangle \nonumber\\
&-\frac{d_i}{4}\left\langle\bb\cdot\left[\left(\bj'\cdot\bnabla'\right)\bb'\right]+\bb'\cdot\left[\left(\bj\cdot\bnabla\right)\bb\right]\right\rangle \label{eq:Tjb}
\end{align}
are the transfer rates corresponding to the channels $\pazocal{BB}$ and $\pazocal{JB}$, respectively. One can associate the net energy flux rates due to the Hall term $\sum_{|\bk'| < k}\widehat{\mathcal{T}_{b\rightarrow b}^{H}}(\bk')$ and $\sum_{|\bk'| < k}\widehat{\mathcal{T}_{j\rightarrow b}^{H}}(\bk')$ to $\sum_{l > n}\sum_{m \leq n} \pazocal{H}^{b,\ b}_{l,\ m}$ and $\sum_{l > n}\sum_{m \leq n} \pazocal{H}^{b,\ j}_{l,\ m}$, respectively if the wavenumber $k$ belongs to the intermediate shell $n$ \citep{Teimurazov_2018}. Whereas the channel $\pazocal{BB}$ consists of a single type of transfer (see Eq.~\eqref{eq:Tbb}), the channel $\pazocal{JB}$ is the combination of various type of transfers (see Eq.~\eqref{eq:Tjb}). The non-monotonic nature of the channel $\pazocal{JB}$ transfer is probably stemming from the interplay between those different type of transfer terms. However, a straightforward conclusion on this remains difficult at the moment.

\subsection{Calculation of mediator specific S2S transfers}
\label{S2S_numerics_fixed_giver}
\begin{figure}
    \centering
    \includegraphics[width=0.9\linewidth]{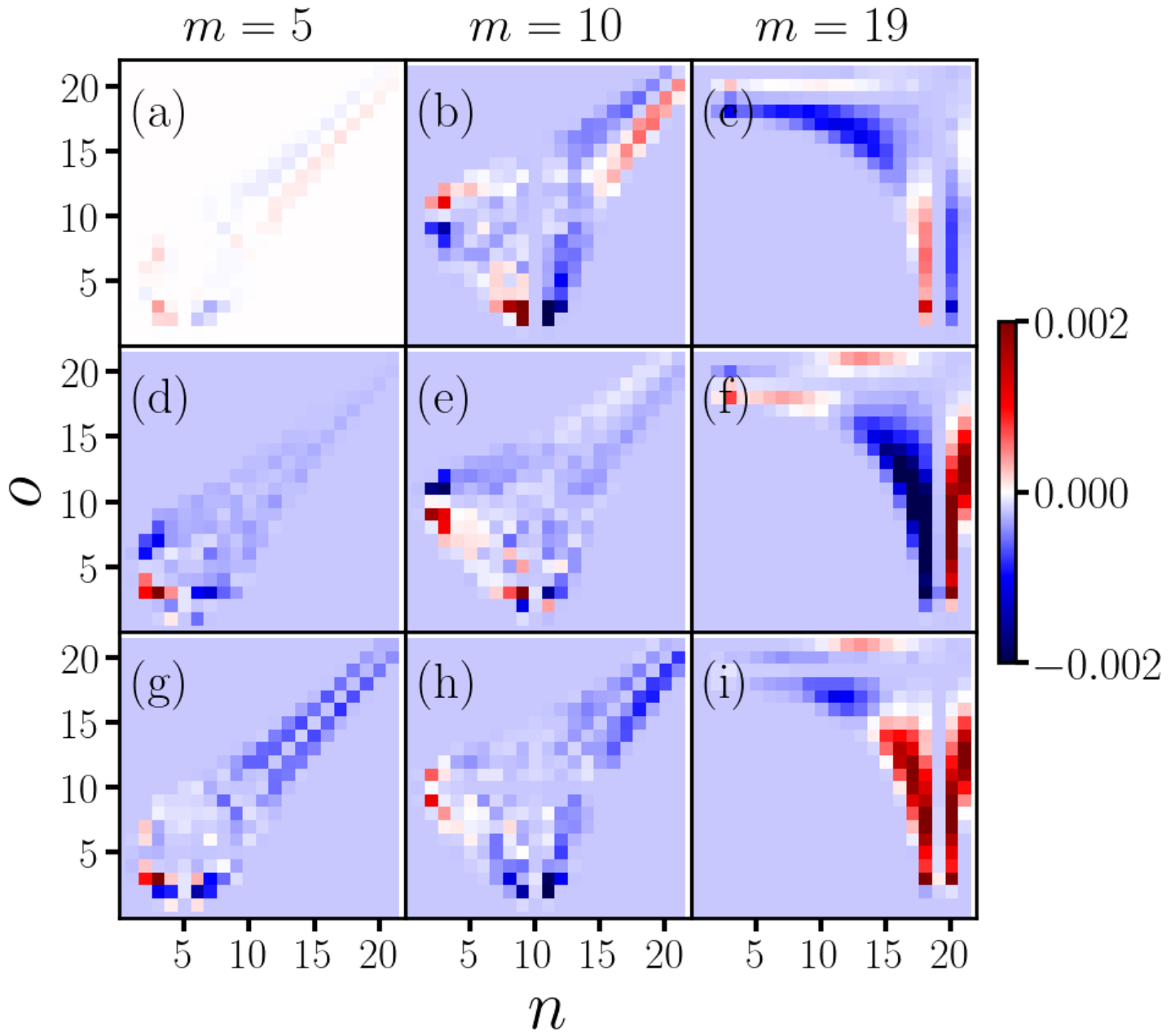}
    \caption{ Top row: $\pazocal{H}^{b,\ b}_{n,\ m,\ o}$ as a function of mediator and receiver shell numbers $(o, n)$ for a giver shell number $m$. Middle row: $\pazocal{H}^{b,\ j}_{n,\ m,\ o}$ as a function of mediator and receiver shell numbers $(o, n)$ for a giver shell number $m$. Bottom row: $|\pazocal{H}^{b,\ j}_{n,\ m,\ o}| - |\pazocal{H}^{b,\ b}_{n,\ m,\ o}|$.  }
    \label{fig:triple_shell_combined_Hall}
\end{figure}
The S2S transfer rates, discussed so far, examine the locality of transfer between a pair of giver-receiver shells $(m,\ n)$ without any information regarding the location of the mediator mode $(q)$. However, the knowledge of $q$ is essential to distinguish between local transfers (comparable giver and receiver wavenumbers) mediated by local ($k\simeq p \simeq q$) and elongated triads $(k \simeq p \gg q)$. For HD and MHD, local transfer via nonlocal triads are found to play important role in the inertial energy cascade \citep{Alexakis_2005b, Mininni_2006, Mininni_2008, Cho_2010}. In order to investigate the same due to the Hall term, we calculate the energy transfer rates as a function of $m,\ n$ and mediator shell $o$:
\textcolor{black}{
\begin{align}
\pazocal{H}^{b,\ b}_{n,\ m,\ o}&=-d_i\sum_{|\bk|\in n}\sum_{|\bq|\in o}\sum_{|\bp|\in m}^{\Delta}\Im\left[\left(\bp\cdot\bj_{\bq}\right)\left(\bb_{\bp}\cdot\bb_{\bk}\right)\right],\label{S2S_bb_triplet}\\
\pazocal{H}^{b,\ j}_{n,\ m,\ o}&=d_i\sum_{|\bk|\in n}\sum_{|\bq|\in o}\sum_{|\bp|\in m}^{\Delta}\Im\left[\left(\bp\cdot\bb_{\bq}\right)\left(\bj_{\bp}\cdot\bb_{\bk}\right)\right].\label{S2S_bj_triplet}
\end{align}}

In top and middle rows of Fig.~\ref{fig:triple_shell_combined_Hall}, we plot $\pazocal{H}^{b,\ b}_{n,\ m,\ o}$ and $\pazocal{H}^{b,\ j}_{n,\ m,\ o}$ for $m = 5,\ 10$ and $19$, respectively. Due to the logarithmic binning in the inertial range, these choices of giver shells approximately corresponds to wavenumbers $\approx 10,\ 20,\ 70$ which captures three regions of $kd_i < 1$, $kd_i \sim 1$ and $kd_i > 1$, respectively (see Table~\ref{tab:shell_indices}). For $\pazocal{H}^{b,\ b}_{n,\ m,\ o}$, regions with weak local transfer is observed for $m = 5$ and close to $o = 5$, corresponding to local triads where $k\simeq p\simeq q$ (see bottom left of Fig.~\ref{fig:triple_shell_combined_Hall}a). Note that for a fixed $(m,\ n)$, we define the triads to be local if $|o-m| \leq 2$ or $|o-n| \leq 2$ and nonlocal otherwise. Another region with a weak but non-zero nonlocal transfer can be identified for $(n,\ o) \gg m$ which naturally associate highly nonlocal triads with $k \simeq q \gg p$ (from center to top right of Fig.~\ref{fig:triple_shell_combined_Hall}a). For $m = 10$, the most intense region of local transfer occurs for $o = 2$ and $3$ associating nonlocal triads with $k \simeq p \gg q$ (deep red and blue squares at the bottom of Fig.~\ref{fig:triple_shell_combined_Hall}b). In this case, two regions of substantial nonlocal transfers are also evident: (i) $m\simeq o \gg n$ corresponding to triads $p \simeq q\gg k$ (blue and red squares at the center left) and (ii) $n\simeq o \gg m$ corresponding to triads $k\simeq q\gg p$ (elongated region at top right). Finally, for $ m = 19$, a similar region of moderate local transfers through nonlocal triads $k\simeq p\gg q$ is observed up to $o \approx 12$ (moderate blue and red squares at bottom right of Fig.~\ref{fig:triple_shell_combined_Hall}c). Beyond this, a horizontal spread towards smaller $n$ values emerges, indicating nonlocal energy transfer involving nonlocal triads where $p\simeq q\gg k$.

For $m = 5$, unlike the previous case, local transfers of $\pazocal{H}^{b,\ j}_{n,\ m,\ o}$ with local triads become significant close to $o = 5$. As with $\pazocal{H}^{b,\ b}_{n,\ m,\ o}$, the nonlocal transfer for $n\simeq o \gg m$ remains extremely weak for this channel (see Fig.~\ref{fig:triple_shell_combined_Hall}d). For $m=10$, similar regions of non-negligible local transfer via nonlocal triads $k\simeq p\gg q$ and nonlocal transfer with nonlocal triads $p\simeq q\gg k$ (dark red and blue squares in the center left of Fig.~\ref{fig:triple_shell_combined_Hall}e) are also observed. However, in contrast to $\pazocal{H}^{b,\ b}_{n,\ m,\ o}$, no substantial nonlocal transfer is evident for the $k\simeq q\gg p$ triads (elongated top right section). For $m = 19$, a region of strong local transfer mediated by nonlocal triads persists up to $o\approx 12$. Beyond which a horizontal spread of diminishing intensity towards smaller $n$ values is observed, suggesting nonlocal energy transfer through triads with $p\simeq q\gg k$ (see Fig.~\ref{fig:triple_shell_combined_Hall}f).

Finally, to compare relative strengths of $\pazocal{H}^{b,\ b}_{n,\ m,\ o}$ and $\pazocal{H}^{b,\ j}_{n,\ m,\ o}$,
we compute the difference $|\pazocal{H}^{b,\ j}_{n,\ m,\ o}| - |\pazocal{H}^{b,\ b}_{n,\ m,\ o}|$ for $m = 5,\ 10$ and $19$ (Fig.~\ref{fig:triple_shell_combined_Hall} bottom row). For each $(o,\ n)$ pair, positive values indicate the dominance of $|\pazocal{H}^{b,\ j}_{n,\ m,\ o}|$ whereas negative values indicate dominance of $|\pazocal{H}^{b,\ b}_{n,\ m,\ o}|$. As it evident, for $m = 5$, the strength of the local transfers with local triads are comparable for the two channels despite a slight dominance of $|\pazocal{H}^{b,\ b}_{n,\ m,\ o}|$  in the region governed by nonlocal transfers (elongated top right section of Fig.~\ref{fig:triple_shell_combined_Hall}g). For $m = 10$, $|\pazocal{H}^{b,\ b}_{n,\ m,\ o}|$ is stronger for regions of local transfer via nonlocal triads (bottom middle of Fig.~\ref{fig:triple_shell_combined_Hall}h) as well as in the nonlocal transfer regime via nonlocal triads $k\simeq q\gg p$ (see elongated top right section). However, $|\pazocal{H}^{b,\ j}_{n,\ m,\ o}|$ takes over in the nonlocal transfer regime associating $p\simeq q\gg k$ triads (see left middle section). Finally, for $m = 19$, $|\pazocal{H}^{b,\ j}_{n,\ m,\ o}|$ clearly dominate in the region of local transfer through nonlocal triads (dark red squares at bottom right of Fig.~\ref{fig:triple_shell_combined_Hall}i). However, $|\pazocal{H}^{b,\ b}_{n,\ m,\ o}|$ slightly takes over $|\pazocal{H}^{b,\ j}_{n,\ m,\ o}|$ in the nonlocal energy transfer regime involving triads with $p\simeq q\gg k$ (see moderate blue squares at center top).

\subsection{Quantitative estimation using power laws}
\begin{figure}
\centering
    \includegraphics[width=\columnwidth]{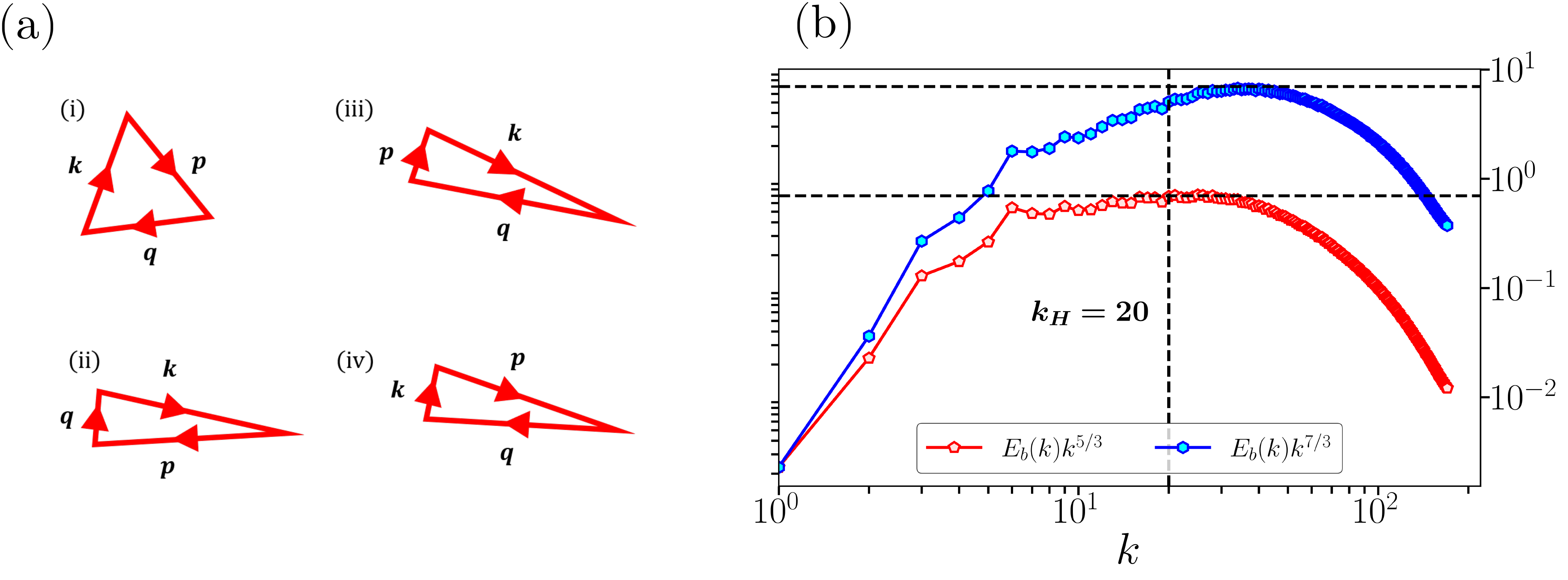}
    \caption{(a) Schematic diagram of contributing triads involved in local and nonlocal transfers and compensated spectra for (b) magnetic energy as a function of wavenumber $k$.}
    \label{triads_magnetic_current_spectrum}
\end{figure}
In the preceding discussion, we have identified different types of triads that contribute to the local and nonlocal transfer of magnetic energy due to the Hall term. These are broadly classified into four categories: (i) $k\simeq p\simeq q$, (ii) $k\simeq p\gg q$, (iii) $k\simeq q \gg p$ and (iv) $p\simeq q \gg k$ (see Fig.~\ref{triads_magnetic_current_spectrum}a). Triads of type (i) and (ii) are primarily involved in local transfers whereas types (iii) and (iv) participate in nonlocal transfers. For a given triad $(\bk, \bp, \bq)$, we can define the ratio 
\begin{equation}
  \pazocal{R} = \frac{|\pazocal{H}^{b,\ b}_{n,\ m,\ o}|}{|\pazocal{H}^{b,\ j}_{n,\ m,\ o}|} \approx \frac{d_i p j_q b_p b_k}{d_i p b_q j_p b_k} = \frac{j_q b_p}{b_q j_p} ,\label{ratio_two_S2S}
\end{equation}
where $\pazocal{R}$ can be taken as a measure of relative strength of the two channels. Assuming a power-law spectra for magnetic energy, one can write $E_b (k)\sim b_k^2 \sim k^{-\alpha} \implies b_k \sim k^{-\alpha/2}$ where $\alpha = 5/3$ and $7/3$ for $k d_i \leq 1$ and $k d_i > 1$, respectively (see Fig.~\ref{triads_magnetic_current_spectrum}b). Due to the algebraic relation $\bj_\bk = i (\bk\times\bb_\bk) \implies j_k^2 = k^2 b_k^2$, leading to $j_k \sim k^{(1-\alpha/2)}$. Using this in the expression \eqref{ratio_two_S2S}, we get $\pazocal{R} = q/p$ irrespective of the wavenumber range. The local triads of type (i) lead to $\pazocal{R} \sim 1$. Interestingly, this is similar to the patterns observed in the S2S transfer rates in Fig.~\ref{fig:triple_shell_combined_Hall} with $m = 10$ and $19$. For $m = 5$, the agreement is weaker which may be due to the fact that the wavenumber region containing triads of type (i) is not completely free from the external forcing effect. In contrast, local transfers via nonlocal triads of type (ii) result in $\pazocal{R} \ll 1$, indicating stronger transfer in the channel $\pazocal{JB}$. This is also similar to the nature of S2S transfer rates in Figs.~\ref{fig:triple_shell_combined_Hall}c, f and i. Triads of type (iii) give $\pazocal{R} \gg 1$ leading to comparatively stronger transfer in the channel $\pazocal{BB}$, consistent with observed enhanced values seen in the top right regions of all the sub-figures of Fig.~\ref{fig:triple_shell_combined_Hall} with $m = 5$ and $10$. Finally, type (iv) triads give $\pazocal{R} \sim 1$, similar to the patterns seen in the top left regions of Figs.~\ref{fig:triple_shell_combined_Hall}c, f and i. Apart from the aforementioned triad configurations, another type of nonlocal triad with $k \simeq p \ll q$ is also possible which leads to $\pazocal{R}\gg 1$. However, due to the cosine rule mentioned earlier, such triads are practically small in number and thus contribute negligibly, as manifested by the negligible transfer rates observed in Fig.~\ref{fig:triple_shell_combined_Hall} for $m = 5$ (top left) and $10$ (top middle). For a better agreement between numerical observations and theoretical results we need a more rigorous analytical framework combined with high-resolution simulations.

\subsection{Effect of the Hall wavenumber}
\label{effect_kH}

\begin{figure}
    \centering
    \includegraphics[width=\columnwidth]{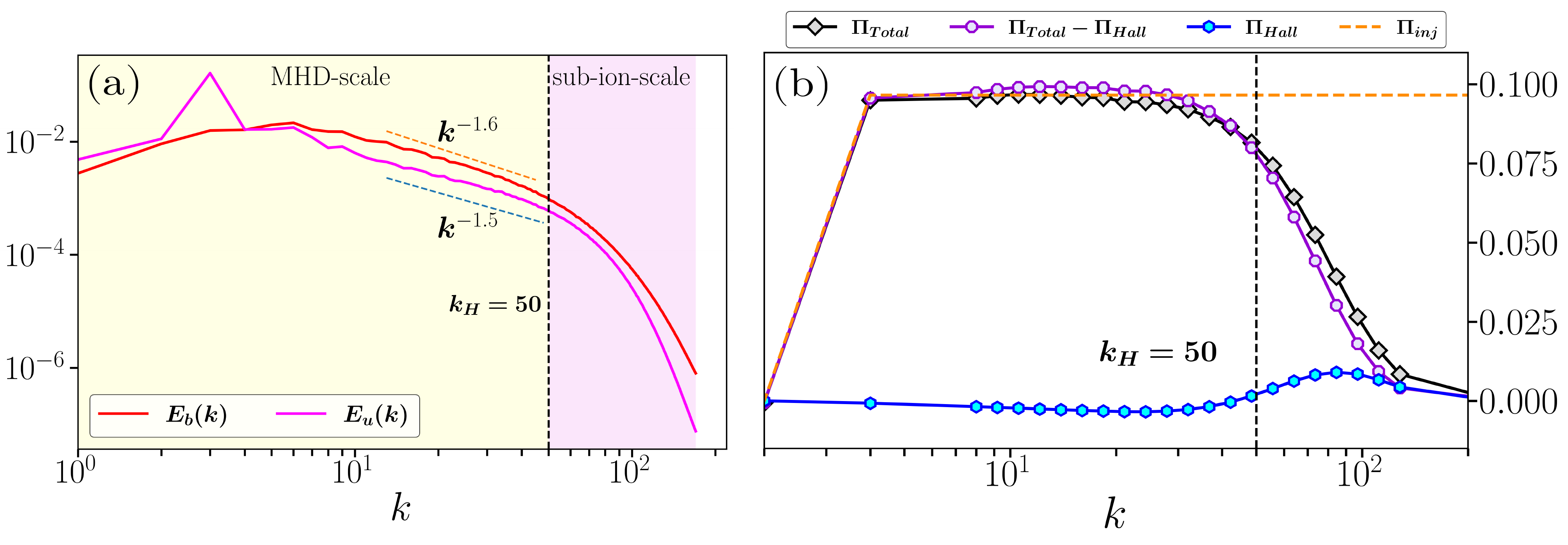}
    \caption{(a) Kinetic and magnetic energy spectra and (b) total energy flux rate, with and without the Hall contribution, as a function of $k$ for the simulation with $k_H = 50$.}
    \label{fig:spectra_flux_kH50}
\end{figure}
\begin{figure}
    \centering
    \includegraphics[width=0.7\columnwidth]{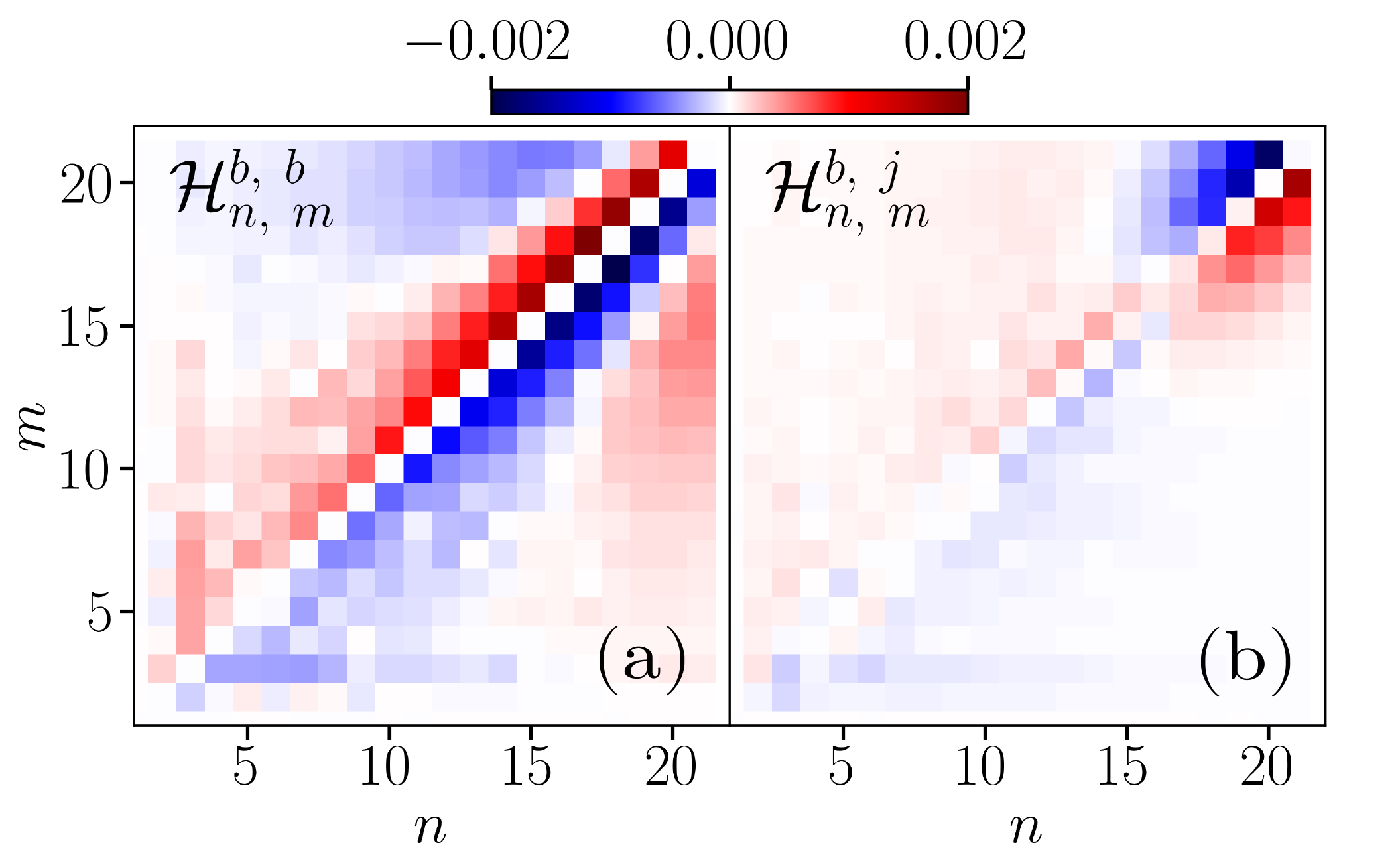}
    \caption{Shell-to-shell transfer rates for the (a) $\pazocal{BB}$ and (b) $\pazocal{JB}$ channels, respectively for the run with $k_H = 50$. }
    \label{fig:S2S_kH50}
\end{figure}

To assess the effect of the Hall wavenumber on the observed results, we run an additional simulation with parameters  identical to those listed in Table \ref{tab:parameters}, except for $d_i = 0.02$, corresponding to $k_H \approx 50$. In the new simulation, we have found an extended inertial range for MHD associating kinetic and magnetic energy spectral exponents of $-1.5$ and $-1.6$, respectively (Fig.~\ref{fig:spectra_flux_kH50}a). This supports our interpretation that the previously observed deviations in the $k_H \approx 20$ run are primarily due to the limited separation between $k_f$ and $k_H$. Despite a weaker $\Pi_{Hall}$ for the region $k>k_H$, here we also observe a range of intermediate wavenumbers where $\Pi_{Total}$ closely follows $\Pi_{inj}$ thus confirming a Kolmogorov-like universal regime for the total energy cascade (Fig.~\ref{fig:spectra_flux_kH50}b). Finally, to examine the effect of $k_H$ on transfer rates, we plot S2S transfer rates for the $\pazocal{BB}$ and $\pazocal{JB}$ channels in Figs.~\ref{fig:S2S_kH50}a and b, respectively. Similar to Fig.~\ref{S2S_HALL_TWO_CHANNELS}a, the channel $\pazocal{BB}$ shows a consistent local inverse transfer for all wavenumbers while a direct nonlocal transfer is observed at small scales. For the channel $\pazocal{JB}$, similar to Fig.~\ref{S2S_HALL_TWO_CHANNELS}b, weak local and nonlocal inverse transfers are observed upto to a certain shell region beyond which it becomes much more enhanced and correspond to a direct magnetic energy transfer. This transition wavenumber (or the shell number) is now shifted towards a higher value with respect to the previous simulation. This analysis clearly shows that the main conclusion of the current work remain effectively unchanged even with the new Hall wavenumber.

\section{Summary and conclusion}
\label{summary}
In this work, we have numerically explored the locality of energy transfer in three-dimensional HMHD turbulence. Noting that the energy is conserved per triad in both the transfer terms associated with $d_i (\bj\cdot\bnabla)\bb$ and $-d_i(\bb\cdot\bnabla)\bj$, we have decomposed the energy transfer via the channels $\pazocal{BB}$ and $\pazocal{JB}$, respectively and checked the locality of the energy transfer in each channel. The total energy flux $\Pi_{Total}$ is found to be constant across the inertial range, confirming a steady Kolmogorov cascade of energy. The kinetic and total magnetic (MHD + Hall) fluxes are also found to be approximately constant thus implying a theoretical description using an almost decoupled cascade for kinetic and magnetic energies is possible in some cases, as also seen in previous studies \citep{Bian_2019}. For $kd_i < 1$, the Hall flux $\Pi_{Hall}$ is slightly negative whereas it becomes positive for $kd_i >1$ and then gradually exceeds $\Pi_u$ and $\Pi_b$.

\begin{figure}
    \centering
    \includegraphics[width=0.8\columnwidth]{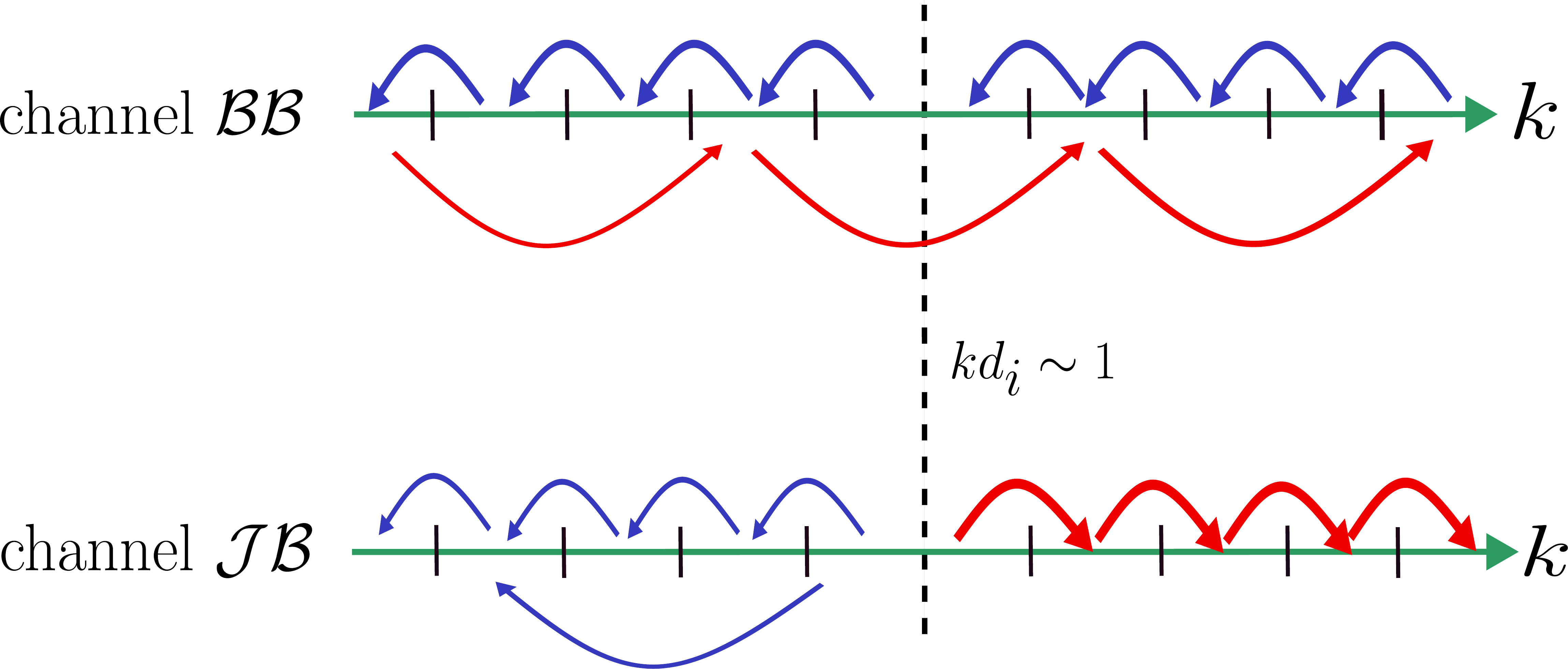}
    \caption{Schematic diagram summarizing the nature of transfers for the two channels of the Hall term. {The red and blue colors indicate the forward and inverse transfers, respectively whereas the thickness of the arrows represent the intensity of the transfers.}}
    \label{fig:two_channel_schematic}
\end{figure}

By computing S2S transfer rates between a pair of giver-receiver shells $(m, n)$, we find that the channel $\pazocal{BB}$ exhibits a strongly local inverse transfer of magnetic energy for all the wavenumbers. Additionally, this channel shows a nonlocal direct transfer that increases at large wavenumbers. On the other hand, both the local and nonlocal inverse transfers of the channel $\pazocal{JB}$ remain relatively weak up to shell number 15 (approximately corresponding to the wavenumber where the slope of magnetic energy spectra changes from $k^{-5/3}$ to $k^{-7/3}$). Beyond that, the transfer becomes positive, accompanied by an increase in intensity of local transfers (see Figs.~\ref{S2S_HALL_TWO_CHANNELS} and \ref{Two_channel_local_flux}, also Fig.~\ref{fig:two_channel_schematic} for a schematic diagram). \textcolor{black}{To convincingly show this, we further showed S2S transfer rates for six receiver shells $n = 5, 9, 11, 13, 15$ and $19$ where the Hall wavenumber is situated between shell numbers $9$ and $11$. The nature of the transfer also indicates that the local transfer for shell $n$ is mainly coming from shells $n\pm 2$.  }

To further examine the nature of fundamental triad interactions, we computed S2S transfer rates for a fixed giver shell $m$ while varying the mediator and receiver shells $o$ and $n$, respectively. In particular, we have chosen $m = 5, 10$ and $19$ to probe the nature of triad interactions in the wavenumber region $kd_i <1$, $kd_i \sim 1$ and $kd_i > 1$, respectively. For $m=5$, both in $\pazocal{H}^{b,\ b}_{n,\ m,\ o}$ and $\pazocal{H}^{b,\ j}_{n,\ m,\ o}$ local transfers via local triads are predominant (see Figs.~\ref{fig:triple_shell_combined_Hall}a, d and g). For $m=10$, similar local transfers via local triads for both the channels decrease whereas the local and nonlocal transfers through nonlocal triads start to develop (see Figs.~\ref{fig:triple_shell_combined_Hall}b, e and h). 
This is further confirmed for $m = 19$ where strong local transfer through nonlocal transfer is observed for both the channels (see Figs.~\ref{fig:triple_shell_combined_Hall}c, f and i). Nonlocal transfers involve two distinct triad configurations $k \simeq q \gg p$ and $p \simeq q \gg k$ (see types (iii) and (iv) in Fig.~\ref{triads_magnetic_current_spectrum}a). Whereas $\pazocal{H}^{b,\ b}_{n,\ m,\ o}$ is predominantly made up of the first kind of non-local triads, leading to enhanced nonlocal transfer at large wavenumbers,  the second type of nonlocal triads are largely present in $\pazocal{H}^{b,\ j}_{n,\ m,\ o}$ at moderate wavenumbers. To explain the different transfer terms in HMHD turbulence, we use power-law assumptions for spectral fields  ($b_k\sim k^{-\alpha/2}$ and $j_k\sim k^{-\beta/2}$). For small wavenumbers ($kd_i< 1$), the model indicates comparable contribution from both the channels whereas for large wavenumbers ($kd_i >1$), the transfer through the channel $\pazocal{JB}$ dominates, thus globally validating our results obtained from the simulation. 

The current work gives a detailed picture of fundamental triadic interactions responsible for nonlocality and backscatter of energy as reported in earlier studies \citep{Mininni_2007, Gomez_2010}. Our results demonstrate that nonlocal triads play a significant role both in the local and nonlocal transfers via the channels $\pazocal{BB}$ and $\pazocal{JB}$. The existing shell model, as the one proposed by \citet{Galtier_2007}, is developed assuming local transfers through local triads only. Our findings mandate a revisit of the same to include additional triadic interactions. Despite its simplicity, HMHD is often used to study the sub-ion scale turbulence in solar wind and magnetospheric plasmas in the vicinity of ion-scale. With the advent of high resolution plasma measurements from multi-spacecraft missions like MMS, it is now possible to directly measure the heating rate associated with different channels of the Hall term \citep{Ferrand_2019}. Our work may be interesting to understand the relative importance between the heating from local and nonlocal interactions at different scales. In a recent work, turbulent relaxation in fluids and plasmas was shown to be explained by principle of vanishing nonlinear transfer (PVNLT) where the relaxed states are associated with zero average nonlinear transfer across inertial scales \citep{Banerjee_2023a, Pan_2024, Sasmal_2025a}. The present study can be useful to understand how the local and nonlocal transfers are affected as a result of such relaxation. The eddy-damped quasi normal Markovian (EDQNM) approximation has been shown to replicate observed locality of energy transfer both in HD and MHD turbulence \citep{Orszag_1970, Pouquet_1976}. A similar investigation for HMHD turbulence to compare the observed results remains a subject for future study. \textcolor{black}{ In addition to the energy, similar investigations can be carried out for magnetic and generalized helicity cascades, which are two other quadratic inviscid invariants of HMHD flow \citep{Yoshida_2002, Banerjee_2016a, Hu_2025}. Beyond HMHD, the current framework may also be extended to the fully developed turbulence of other complex flows including electron MHD and two-fluid plasmas \citep{Celani_1998a, Banerjee_2020, Banerjee_2025}, ferrofluids \citep{Mouraya_2019, Mouraya_2024}, binary fluids \citep{Pan_2022, Pan_2025} \textit{etc}. to obtain a clear anatomy of the locality of the non-linear interactions across the inertial scales.}

\section*{Acknowledgments}
The authors acknowledge one of the anonymous referees for the valuable suggestion to verify the robustness of our results for a different Hall wavenumber. The authors express sincere gratitude to the Editor, Dr. Thierry Passot, for his insightful comments and valuable suggestions, particularly regarding the calculation of the flux rates.   A.H. and S.B. acknowledge the financial support from STC-ISRO grant (STC/PHY/2023664O). The simulation code is developed by A.H. following the parallelization schemes in  \citet{Mortensen_2016}. The simulation is performed using the support and resources provided by PARAM Sanganak under the National Supercomputing Mission, Government of India at the Indian Institute of Technology, Kanpur.

\section*{Declaration of interests}
The authors report no conflict of interest.

\appendix
\section{}\label{appA}
The volume of the $n$-th shell for linear and logarithmic binning are given by
\begin{equation}
\Delta V_{lin} = \frac{4}{3}\pi\left[k_n^3 - (k_n - 1)^3\right]\;\;\; \text{and}\;\;\; \Delta V_{log} = \frac{4}{3}\pi\left[k_n^3 - \left(\frac{k_n}{\beta}\right)^3 \right],
\end{equation}
respectively where $k_n$ is the outer radius of the $n$-th shell and $\beta = k_n/k_{n-1}$. Simplifying and assuming $1/k_n = \epsilon\ll 1$, the above two expressions one can write 
\begin{equation}
\Delta V_{lin} = \frac{4}{3}\pi k_n^3\left[1 -\left(1-\epsilon\right)^3 \right]\simeq \frac{4}{3}\pi k_n^3 (3\epsilon) \;\;\; \text{and}\;\;\; \Delta V_{log} = \frac{4}{3}\pi k_n^3 \left(1-\frac{1}{\beta^3}\right). 
\end{equation}
For $\beta = 1.15$, it is easy to see that $\Delta V_{log} > \Delta V_{lin}$ for $k_n>8$, where the logarithmic binning accommodates a large number of high wavenumber Fourier modes.

\section{}\label{appB}
The net transfer rate from $\bm{y}$-to-$\bm{x}$ across a sphere of radius $k_0$ field can be written as 
\begin{equation}
    \Pi^{xy}(k_0)
=
-\sum_{|\bk|\leq k_0}\sum_{|\bp|>  k_0}^{\Delta}
S^{xy}(\bm{k}|\bm{p}|\bm{q})
-
\sum_{|\bk|\leq k_0}\sum_{|\bp|\leq  k_0}^{\Delta}
S^{xy}(\bm{k}|\bm{p}|\bm{q}).\label{y2x}
\end{equation}
The back transfer from  $\bm{x}$-to-$\bm{y}$ field, if exists, can be written as
\begin{equation}
    \Pi^{yx}(k_0)
=
-\sum_{|\bk|\leq k_0}\sum_{|\bp|>  k_0}^{\Delta}
S^{yx}(\bm{k}|\bm{p}|\bm{q})
-
\sum_{|\bk|\leq k_0}\sum_{|\bp|\leq  k_0}^{\Delta}
S^{yx}(\bm{k}|\bm{p}|\bm{q}).\label{x2y}
\end{equation}
The first term on the r.h.s of \eqref{y2x} and \eqref{x2y} cannot permit any swapping between $\bk$ and $\bp$ as $|\bk|$ and $|\bp|$ can take mutually exclusive values. However, the second terms permit a $\bk$-$\bp$ swapping and hence one can write 
\begin{equation}
-
\sum_{|\bk|\leq k_0}\sum_{|\bp|\leq  k_0}^{\Delta}
S^{yx}(\bm{k}|\bm{p}|\bm{q}) = -
\sum_{|\bp|\leq k_0}\sum_{|\bk|\leq  k_0}^{\Delta}
S^{yx}(\bm{p}|\bm{k}|\bm{q}).
\end{equation}
Again, due to the giver-receiver anti-symmetry one may write 
\begin{equation}
-
\sum_{|\bp|\leq k_0}\sum_{|\bk|\leq  k_0}^{\Delta}
S^{yx}(\bm{p}|\bm{k}|\bm{q}) = 
\sum_{|\bp|\leq k_0}\sum_{|\bk|\leq  k_0}^{\Delta}
S^{xy}(\bm{k}|\bm{p}|\bm{q}),  
\end{equation}
thus leading to the mutual cancellation of the last two terms on the r.h.s of \eqref{y2x} and \eqref{x2y}. As mentioned earlier, in HMHD, the back transfer of the channel $\pazocal{JB}$ does not appear in the energy transfer. It is therefore necessary to include the contribution from $\sum_{|\bp|\leq k_0}\sum_{|\bk|\leq  k_0}^{\Delta}
S^{bj}(\bm{k}|\bm{p}|\bm{q})$ to calculate the correct flux rate.


\bibliographystyle{jpp}
\bibliography{main}

\end{document}